\documentclass[useAMS,usenatbib]{mn2e}
\usepackage{amsmath,amstext,amsgen,amsbsy,amsopn,amsfonts,theorem}
\usepackage{natbib}
\pdfoutput=1

\usepackage{graphicx}
\usepackage{xspace}
\usepackage{amsmath}
\usepackage{hyperref}
\usepackage{framed}
\usepackage{txfonts}
\usepackage{epstopdf}
\usepackage{gensymb}

\usepackage{booktabs}

\usepackage[normalem]{ulem}

\makeatletter
\def\mr@ignsp#1 {\ifx\:#1\@empty\else #1\expandafter\mr@ignsp\fi}%
\newcommand{\multiref}[1]{\begingroup%\let\protect\string%
\xdef\mr@no@sparg{\expandafter\mr@ignsp#1 \: }%
\def\mr@comma{}%
\@for\mr@refs:=\mr@no@sparg\do{\mr@comma\def\mr@comma{,}\ref{\mr@refs}}%
\endgroup}
\makeatother

\newcommand{\hypref}[2]{\ifx\href\asklfhas #2\else\href{#1}{#2}\fi}
\newcommand{\Secref}[1]{Section~\multiref{#1}}
\newcommand{\secref}[1]{Sec.~\multiref{#1}}

\newcommand{\Tabref}[1]{Table~\multiref{#1}}

\newcommand{\Figref}[1]{Figure~\multiref{#1}}
\newcommand{\figref}[1]{Fig.~\multiref{#1}}
\renewcommand{\eqref}[1]{(\multiref{#1})}

\newcommand{\eq}[1]{\begin{align}#1\end{align}}

\newcommand{\nln}{\nonumber\\}

\title[Properties of Dark Subhaloes from Gaps in Tidal Streams]{Properties of Dark Subhaloes from Gaps in Tidal Streams}

\author[D. Erkal, V. Belokurov]
  {Denis Erkal$^1$\thanks{derkal@ast.cam.ac.uk}, Vasily Belokurov$^1$ \\
  $^1$Institute of Astronomy, Madingley Road, Cambridge, CB3 0HA, UK}

\begin{document}

\label{firstpage}

\maketitle

\begin{abstract}
Cold or Warm, the Dark Matter substructure spectrum must extend to
objects with masses as low as $10^7 M_\odot$, according to the most
recent Lyman-$\alpha$ measurements. Around a Milky Way-like galaxy,
more than a thousand of these subhaloes will not be able to form stars
but are dense enough to survive even deep down in the potential well
of their host. There, within the stellar halo, these dark pellets will
bombard tidal streams as they travel around the Galaxy, causing small
but recognizable damage to the stream density distribution. The
detection and characterization of these stream ruptures will allow us
to constrain the details of the subhalo-stream interaction. In this
work, for the first time, we will demonstrate how the properties of a
subhalo, most importantly its mass and size, can be reliably inferred
from the gap it produces in a tidal stream. For a range of realistic
observational setups, mimicking e.g. SDSS, DES, Gaia and LSST data, we
find that it is possible to measure the {\it complete set} of
properties (including the phase-space coordinates during the flyby) of
dark perturbers with $M>10^7 M_\odot$, up to a 1d degeneracy between
the mass and velocity.
\end{abstract}


\section{Introduction}

Embarrassingly few observational techniques exist with which to
reliably gauge the presence of low-mass Dark Matter (DM) clumps in the
Universe around us. Even though swarms of completely dark subhaloes
are predicted to orbit galaxies like our own
\citep{Diemand2008,springel_et_al_2008}, there is still no proven
tactic to detect them. Until recently, the only acclaimed direct probe
has been strong gravitational lensing \citep[see e.g.][]{Dalal2002,
  Hezaveh2014}. However, such measurements are statistical in nature:
only the total mass fraction locked in small sub-structure can be
typically estimated. The largest of the small subhaloes, those with
masses between $10^8 M_\odot$ and $10^9 M_\odot$ can perhaps be
identified individually \citep[see e.g.][]{Vegetti2012,
  Hezaveh2013}. Yet, there is little hope to prove that they do not
possess a substantial baryonic component.

In the Milky Way, arguably, the frontier of low-luminosity low-mass
substructure detection was breached long ago. A population of
DM-dominated dwarf satellites is known to surround the Galaxy
\citep[see e.g.][]{Aaronson1983,Mateo1998,Tolstoy2009}. For every
solar mass of hydrogen, these dwarfs contain hundreds of solar masses
of Dark Matter \citep[see e.g.][]{Walker2009}. Granted, the total
masses of the subhaloes the dwarf spheroidals inhibit are not
currently known. The initial mass of the faintest of them can be as
high as $10^9 M_\odot$ \citep[see e.g.][]{Wheeler_2015}. However, it
is not implausible that they could have started with as little as
$10^7 M_\odot$ of DM \citep{Bland-Hawthorn2015}. By redshift zero,
their dark masses have probably been depleted further while their
luminosities are at the level barely seeable at distances far smaller
than cosmological \citep[see e.g.][]{Belokurov2013}.

Between $10^7 M_\odot$ and $10^9 M_\odot$, there ought to be roughly
2000 DM subhaloes in a $10^{12} M_\odot$ host if DM were cold
\citep{springel_et_al_2008}. As of today, the census of the Milky
Way's dwarf population is incomplete, but the total is predicted not
to exceed several hundred satellites \citep[see][]{Tollerud_2008,
  Koposov_2008, Koposov_2009}. It is clear that more than half of the
substructure in this mass regime must remain invisible. Recently, Warm
Dark Matter has been invoked as a silver bullet to make the problems
associated with low-mass substructure go away \citep[see
  e.g.][]{Lovell_2012, Lovell_2014}. However, in line with the
up-to-date Ly $\alpha$ measurements, if the DM were warm, its mass
ought to be greater than 3.3 keV at the 2$\sigma$ level \citep[see
  e.g.][]{Viel_2013}. This, in turn, leaves very little wriggle room
to modify the substructure mass function above $v_{\rm max} \approx 6$
km s$^{-1}$ \citep[see][]{Schneider_2014}.

The next goal, therefore, is to discover subhaloes totally devoid of
stars. The most promising avenue is the detection and characterization
of the perturbations DM clumps induce on stellar streams in the Galaxy
\citep[e.g.][]{ibata_et_al_2002,johnston_et_al_2002}. Through a series
of numerical experiments, it has now been established that dark
subhaloes should interact with stellar streams sufficiently
frequently, leaving small but observable signatures in their wake
\cite[e.g.][]{siegal_valluri_2008,carlberg_2009,carlberg_2012,yoon_etal_2011,ngan_carlberg_2014,ngan_carlberg_2015,carlberg_2015}. For
example, during the lifetime of a Pal 5-like stream roughly 5 direct
impacts with subhaloes with $10^7 M_\odot < M < 10^8 M_\odot$ are
expected and 20 impacts with subhaloes with $10^6 M_\odot < M < 10^7
M_\odot$ \citep{yoon_etal_2011}. There are several tens of tidal streams already discovered
in the Milky Way \citep{newberg_carlin} - plenty of potential
carriers of signs of past interactions.

Following the treatment introduced in \citet{carlberg_2013}, we laid
down the basic equations governing the formation of stream
perturbations in \citet[][Paper I]{erkal_belokurov_2015}. Paper I
presents a simple picture of the three phases of stream gap evolution,
and, most importantly, demonstrates that a measurement of the gap density profile alone
is not sufficient to break the strong degeneracy between the
perturber's mass and size, the impact geometry and kinematics, and the time since flyby. However, if complementary
phase-space measurements around the gap were collected, would the
evidence assembled be ample to reconstruct what exactly happened
during the run-in? This is the question we strive to answer in this
Paper.

In this work we will simulate cold streams similar to that of Pal 5
\citep[see e.g.][]{pal5disc}. We will consider the 6d phase-space
structure of the gap and show that using this expanded set of
observables, it is possible to infer the properties of the subhalo's
flyby, up to a 1d degeneracy. This is a degeneracy between the subhalo
mass and the velocity of its flyby which is unavoidable since it is
actually a degeneracy in the velocity kick imparted on the stream
particles during the flyby. We will show that with proper motions and
distances from Gaia, Dark Energy Survey (DES)-like photometry, and moderate resolution
spectroscopy for radial velocities, it will be possible to detect and
characterize dark subhaloes down to $M \sim 10^7 M_\odot$ within 10
kpc. We will also discuss the mass range which can be probed by
upcoming surveys like the Large Synoptic Survey Telescope (LSST) and give general expressions for the mass
sensitivity as a function of the errors in each of the 6d phase-space
coordinates.

Finally, we deliberate on the constraints that can be derived if only
a subset of the observations can be secured. For example, for
sufficiently distant streams it may not be possible to get proper
motions with Gaia. We show that in general one only needs to measure
the following: the shape of the gap on the sky, the density profile of
the gap, and the line-of-sight velocity along the gap. Fortunately,
these are the simplest measurements to make for targets across a large
range of heliocentric distances, so this does not limit our ability to
infer subhalo properties.

This paper is organized as follows. We give the setup and motivation
in \Secref{sec:setup}. In \Secref{sec:observables}, we present
analytic expressions for the 6d structure of the gap. Next, in
\Secref{sec:basic_inference}, we use these results to infer the
properties of a subhalo from a gap in an N-body simulation without
accounting for observational errors. In
\Secref{sec:information_content}, we discuss the information content
of each observable and find the minimal set of observables which are
needed to constrain subhalo properties. In
\Secref{sec:error_inference}, we infer the properties of subhaloes
from gaps in simulations using observational errors of current and
future surveys. Continuing with realistic errors, we present analytic
expressions for the mass range accessible as a function of survey
errors in \Secref{sec:mass_range}. In \Secref{sec:discussion} we
discuss complications which will arise when extending this model to
more realistic streams. Finally, we conclude in \Secref{sec:conclusions}.

\section{Setup and Motivation} \label{sec:setup}

The aim of this work is to show that it is possible to start with a gap in a tidal stream and infer the properties of the subhalo which created it. In order to study this problem, we will consider a simple setup with a stream on a circular orbit which is perturbed by the flyby of a Plummer sphere. This is identical to the setup of Paper I. In that work, we were primarily interested in the density structure of the gap and we showed that a measurement of the stream density profile near the gap does not uniquely constrain the properties of the subhalo which created the gap. In fact, it only provides constraints on three out of the seven parameters which describe the subhalo flyby. Thus, more constraints are needed. The results of Paper I also contain predictions for the entire 6d structure of the gap which we will exploit in this work. With this extra information, we will show that we can constrain the subhalo properties down to a 1d degeneracy between mass and velocity.

One of the main benefits of this simplistic model for gaps is that all of the predictions are analytic. With these expressions for the stream shape, we can predict the subsequent evolution of the stream after a subhalo flyby. For a given gap, the model can then be used to infer the properties of the subhalo which created it. Furthermore, the analytic expressions will allow us to probe the constraining power of each observable and we will discuss what degeneracies remain if only certain observations are possible. Finally, we will use these expressions to give constraints on what mass range is gaugeable given the errors of current and future surveys. Without these analytic predictions, a lot of this would not be possible and we would need to build intuition entirely through individual numerical examples.

\begin{figure}
\centering
\includegraphics[width=0.3\textwidth]{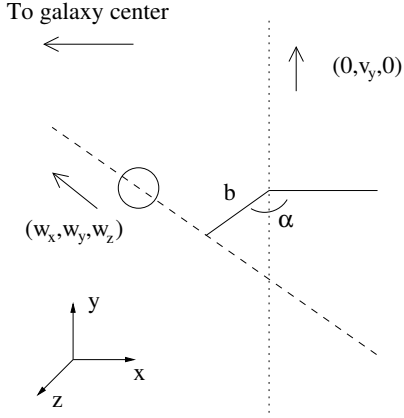}
\caption{Our axis convention looking down on the stream. The dotted line is the stream and the dashed line is the path of the subhalo. The solid lines show the impact
parameter, $b$, and its angle in the $x-z$ plane, $\alpha$. The center of the host potential is to the left and the orbit is counterclockwise. Note that the angle $\alpha$ can be re-expressed in terms of the velocities $w_x$ and $w_z$ since the velocity is perpendicular to the impact parameter at closest approach. This determines the impact parameter up to an overall sign which we include in $b$. We note that this figure is identical to Figure 1 in Paper I. } 
\label{fig:axes}
\end{figure}

\Figref{fig:axes} displays the geometry of the subhalo-stream encounter as used in this work. The stream is on a circular orbit and we choose our coordinates to be centered on the point of closest approach along the stream with $x$ denoting the radial direction, $y$ denoting the direction along the stream, and $z$ denoting the direction perpendicular to the stream plane. This is identical to the setup in Paper I. We assume that the stream is orbiting in a spherical potential which is otherwise unconstrained. The stream is at a radius $r$ with a circular velocity $v_y = \sqrt{r \partial_r \phi}$. This model allows for any impact geometry. We denote the subhalo velocity as $(w_x,w_y,w_z)$ and the stream velocity as $(0,v_y,0)$. At the point of closest approach, the vector between the stream and the subhalo must be perpendicular to the relative velocity $(w_x,w_y-v_y,w_z)$. This relative velocity, along with the impact parameter $b$, defines the subhalo's trajectory up to an overall sign since the subhalo could pass on either side of the stream. We will encode this sign in the impact parameter, $b$. Next we have the mass of the subhalo, $M$, and its scale radius, $r_s$. Finally, we have the time since impact, $t$. For a given gap, we take the impact point to be the current position of the gap center integrated backwards in time by $t$. Thus we have seven model parameters: three velocities, $\vec{w}$, the impact parameter, $b$, the mass $M$, the scale radius, $r_s$, and the time since impact, $t$. Following the convention of Paper I, we break the relative velocity between the stream and subhalo into a velocity along the stream, $w_\parallel = v_y - w_y$, and a perpendicular velocity of $w_\perp = \sqrt{w_x^2+w_z^2}$. The magnitude of the relative velocity is denoted by $w = \sqrt{w_\parallel^2 + w_\perp^2}$.

During the flyby, we assume that the stream is moving along a straight line. This is needed to compute the velocity kicks along the stream. Since the velocity kicks take place over a relatively small distance, several kpc at most for the most massive subhaloes we will consider, this is a reasonable assumption.

\section{New Observables} \label{sec:observables}

The first observable we will consider is the density profile along the stream as in Paper I. In addition, we will now use the analytic model produced in Paper I to make predictions for the 6d phase space structure of the stream after the flyby, giving five new observables. First we have the spatial shape of the stream near the gap which gives two observables: 1) the distance to the stream as a function of the angle on the sky, $x$ vs $\psi$ (where $\psi$ is the angle along the stream) and 2) the shape of the stream on the sky, $z$ vs $\psi$. Next we have the velocity structure of the gap which gives three observables: 3) the radial velocity as a function of the angle on the sky, $\dot{x}$ vs $\psi$, 4) the proper motion along the stream as a function of the angle on the sky, $\dot{y}$ vs $\psi$, and 5) the velocity perpendicular to the orbital plane as a function of the angle on the sky, $\dot{z}$ vs $\psi$. We note that these observables are what an observer in the center of the galaxy would measure. They do not account for the solar position or velocity. We will give analytic predictions of these observables below and then use them to infer the subhalo from several N-body simulations. Note that these predictions can only be expressed parametrically, i.e. to describe the shape of the gap on the sky ($z$ vs $\psi$) at time $t$, we express both $z$ and $\psi$ in terms of the angle from the point of closest approach, $\psi_0$. 

Crucially, the structure of the gap is controlled by the velocity kicks imparted on the stream during the flyby. The kicks in the $x$, $y$, and $z$ direction (radial, along the stream, and perpendicular to the orbital plane), as a function of distance from the point of closest approach were computed analytically for a Plummer sphere perturber in Paper I:

\eq{ \Delta v_x(\psi_0) &= \frac{2GM }{r_0^2 w_\perp^2 w 
} \frac{\Big( b w^2 \frac{w_z}{w_\perp} -  \psi_0 r_0 w_\parallel w_x  \Big)}{\Big(\psi_0^2 + \frac{(b^2+r_s^2) w^2}{r_0^2 w_\perp^2} \Big)} ,  \nln
 \Delta v_y(\psi_0) &= -\frac{2GM \psi_0}{w r_0 \Big( \psi_0^2+  \frac{(b^2+r_s^2) w^2}{r_0^2 w_\perp^2} \Big)} ,\nln 
 \Delta v_z(\psi_0) &= -\frac{2 G M }{r_0^2 w_\perp^2 w } \frac{\Big(b w^2 \frac{w_x}{w_\perp} + \psi_0 r_0 w_\parallel w_z\Big)}{\Big( \psi_0^2 + \frac{(b^2+r_s^2) w^2}{r_0^2 w_\perp^2} \Big)}.\label{eq:deltav} }
These formulae are slightly different than the ones in Paper I since we have re-expressed the kicks in terms of the three components of the flyby velocity instead of two velocity components and an angle. We have also re-expressed the distance from the point of closest approach, $y$, in terms of the angle along the stream, i.e. $y = r_0 \psi_0$. Next, we list the prediction for each of the 6d phase space coordinates.

\subsection{$\psi$}

First we have the angle along the stream. This is given in equation 16 of Paper I:

\eq{ \psi(\psi_0,t) = \psi_0 + \frac{f \psi_0 - g}{\psi_0^2 + B^2}. \label{eq:psi_t}}
where $\psi_0 = \frac{y}{r_0}$ denotes the initial distance between a point along the stream and the point of closest approach, 
\eq{ f = \frac{4-\gamma^2}{\gamma^2} \frac{t}{\tau} - \frac{4
\sin\big(\frac{\gamma v_y t}{r_0} \big)}{\gamma^3} \frac{r_0 }{v_y \tau} 
+ \frac{2 \Big( 1 - \cos\big( \frac{\gamma v_y t}{r_0}\big) \Big)}{\gamma^2}
\frac{w_\parallel w_x}{w_\perp^2} \frac{r_0 }{v_y \tau} , \label{eq:f} }
\eq{ g = \frac{2 \Big( 1 - \cos\big( \frac{\gamma v_y t}{r_0}\big)
\Big)}{\gamma^2} \frac{b w^2 w_z}{r_0 w_\perp^3} \frac{r_0 }{v_y \tau}, \label{eq:g} }
\eq{ B^2 = \frac{b^2 + r_s^2}{r_0^2} \frac{w^2}{w_\perp^2}, \label{eq:B} }
\eq{ \gamma^2 = 3 + \frac{r_0^2}{v_y^2} \partial_r^2 \phi(r_0), }
and the timescale $\tau$ is given by
\eq{ \tau = \frac{w r_0^2}{2 GM} . \label{eq:tau}}
Note that these expressions are slightly different from those in Paper I since we have re-written the angle $\alpha$ in terms of ratios of $w_x$, $w_z$, and $w_\perp$. The density is related to this angle since $\rho \propto | \frac{d\psi}{d\psi_0}|^{-1}$.

\subsection{$x$}

The radial distance of each particle in the stream is computed in equation 8 of Paper I:

\eq{ \Delta x(\psi_0,t) =  \frac{2 r_0 \Delta v_y(\psi_0)}{v_y} \frac{ \Big( 1 - \cos( \frac{\gamma v_y t}{r_0})\Big)}{\gamma^2} + \frac{r_0 \Delta v_x(\psi_0)}{v_y} \frac{\sin(\frac{\gamma v_y t}{r_0})}{\gamma} \label{eq:x},}
where the velocity kicks, $\Delta v_x(\psi_0), \Delta v_y(\psi_0)$, are specified in \eqref{eq:deltav}. 

\subsection{$z$}

The kicks in the $z$ direction tilt the orbital plane of the resulting orbits by an angle of $\Delta v_z/v_y$ in the $\hat{x}$ direction, therefore the offset in stream latitude of the particles is given by

\eq{ \Delta \delta_z (\psi_0,t) = \frac{\Delta v_z (\psi_0)}{v_y} \sin( \frac{v_y t}{r_0}) \label{eq:z},}
where $\Delta \delta_z = \frac{\Delta z}{r_0}$ which denotes the difference in stream latitude from the unperturbed stream track.

\subsection{$\dot{x}$}

The radial velocity can be derived from a time derivative of \eqref{eq:x} which gives

\eq{ \Delta \dot{x}(\psi_0,t) = \frac{2 \Delta v_y(\psi_0)}{\gamma} \sin(\frac{\gamma v_y t}{r_0}) + \Delta v_x(\psi_0)\cos( \frac{\gamma v_y t}{r_0}). \label{eq:dotx}}

\subsection{$\dot{y}$}

The velocity along the stream is given by 

\eq{ \Delta \dot{y}(\psi_0,t) &= \left(r_0 + \Delta x(\psi_0,t)\right) \dot{\theta}(\psi_0,t) - v_y ,}
where $\dot{\theta}(\psi_0,t)$ is the angular velocity along the stream and is given in equation 12 of Paper I and $\Delta x(\psi_0,t)$ is given in \eqref{eq:x}. Plugging in these expressions and taking only the leading order terms we get
\eq{ \Delta \dot{y}(\psi_0,t) = - \Delta v_y(\psi_0) \frac{2 - \gamma^2}{\gamma^2} + 2 \Delta v_y(\psi_0) \frac{\cos( \frac{\gamma v_y t}{r_0}) }{\gamma^2} - \Delta v_x(\psi_0) \frac{\sin (\frac{\gamma v_y t}{r_0} )}{\gamma} . \label{eq:doty} } 
We note that we only consider the leading order terms in $\Delta v_x$ and $\Delta v_y$ in this equation since the analysis in Paper I was done at leading order.

\subsection{$\dot{z}$}

The velocity in the $z$ direction comes from a time derivative of the offset in the $z$ direction, i.e. the time derivative of \eqref{eq:z}:
\eq{ \Delta \dot{z}(\psi_0,t) = \Delta v_z(\psi_0) \cos( \frac{v_y t}{r_0}) \label{eq:dotz} .}
By combining these six results, \eqref{eq:psi_t} and \eqref{eq:x}-\eqref{eq:dotz}, we have a parametric model for the phase space structure of the stream at time $t$ in terms of the initial position along the stream, $\psi_0$. In \Figref{fig:1e8_shape}-\Figref{fig:1e7_shape}, we show a comparison of the model presented in this section against the results of N-body simulations which we will describe in the next section. From these Figures, it is evident that the model can successfully reproduce the structure of the gap.

\section{Inference without errors}\label{sec:basic_inference}

Now that we have introduced the new observables, we will use a suite of N-body simulations to demonstrate that the model works. We will then run a set of MCMCs to show that the model can be used to constrain the properties of the subhalo which created the gap.

\subsection{Simulations} \label{sec:simulations}

To generate the gaps we carry out three N-body simulations with the pure N-body part of \textsc{gadget-3} which is closely related to \textsc{gadget-2} \citep{springel_2005}. 

The setup is similar to the realistic simulations in Paper I. We first generate a stream 
by taking a globular cluster modelled as a King profile with a mass of $10^5 M_\odot$, a core radius of 13 pc, and $w=5$. This progenitor is placed on a circular orbit
with a radius of 10 kpc around an NFW potential \citep{nfw_1997} with $M = 10^{12} M_\odot$, $c=15$  and $R_s = 14.0$ kpc. The circular velocity at 10 kpc is $v_y = 168.2$ km/s. The King profile is represented with $10^6$ particles which have a smoothing length of 1 pc. The cluster is evolved for 5 Gyr and in this time a cold stream with a total length of $\sim 180 \degree$ is produced. We take this stream and place a subhalo on an orbit to impact the center of the leading arm of the stream, which corresponds to a radius of 9.80 kpc, at 150 km/s perpendicular to the orbital plane. We then follow the evolution of the stream for an additional 5 Gyr. We have chosen a simple impact geometry in this example but the analytic model works for any impact geometry.

In our experiments, we use three different subhaloes to showcase the range of masses and scale radii for which we will be able to infer the subhalo properties. These are chosen to roughly lie in the center of the scale radius-mass relation in the public data release of the Via Lactea II simulation \citep{Diemand2008}. First, we have the largest subhalo with $M = 10^8 M_\odot$ with $r_s = 625$ pc. Next, we have an intermediate subhalo with $M = 10^{7.5} M_\odot$ with $r_s = 395.3$ pc. Finally, we have the smallest subhalo with $M=10^7 M_\odot$ and $r_s = 250$ pc. We note that while these subhaloes were chosen in the middle of the scale radius-mass relation from Via Lactea II, we have modelled them with Plummer spheres instead of NFW profiles. As a result, our Plummer spheres reach a maximum circular velocity at a smaller radius than if they were modelled with NFW profiles. If we instead compare the radius at which the maximum circular velocity of our subhaloes is reached to those in Via Lactea II, we find that our subhaloes are still within the scatter but are now closer to the edge of the relation rather than the center.

As we saw in \Secref{sec:observables}, the amplitude of the oscillations induced by the subhalo's passage is governed by the maximum velocity kick induced by the subhalo. For a direct impact, this maximum is proportional to $M/r_s$ which follows from \eqref{eq:deltav}. The properties of these subhaloes are arranged so that the size of the perturbation decreases by a factor of two as we go from the most massive to the intermediate subhalo, and another factor of two as we go from the intermediate to the least massive subhalo. Finally, we have to choose how long after impact to look at each of the gaps. We choose to look at the perturbations when the deflection on the sky ($z$ vs $\psi$) is relatively large at $t=450$ Myr. Note that the different subhaloes will have created different sized gaps by this time. While the gap will grow in time, becoming easier to observe, we choose to look at the gap after a short time where our model is most accurate. At late times, the distribution of energy and angular momenta along the stream can cause the gap to fill in which is not captured by our model. We choose to look at all subhaloes at the same time to make it a fair comparison.

\begin{figure}
\centering
\includegraphics[width=0.5\textwidth]{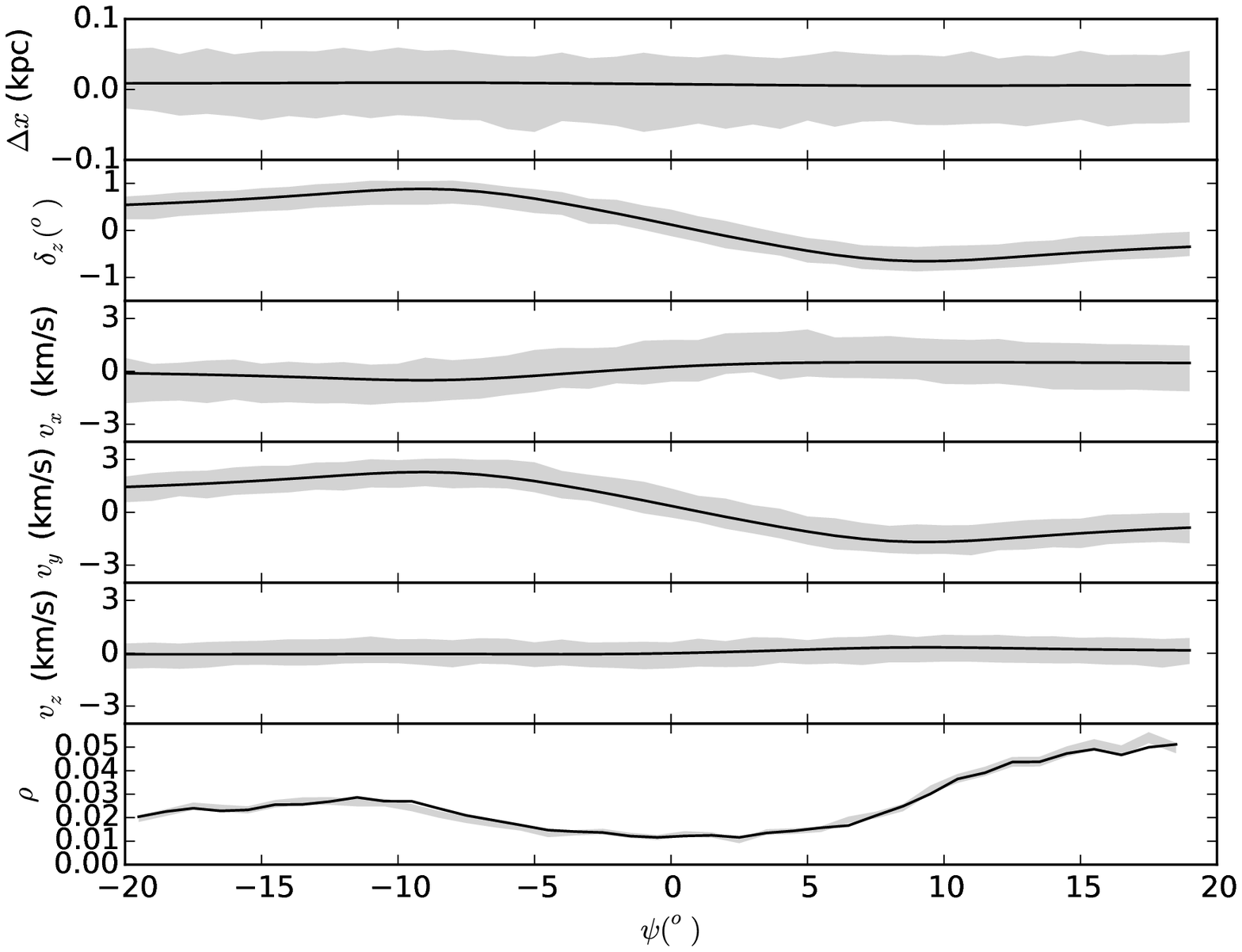}
\caption{Gap observables for a flyby of a Plummer sphere with a mass of $10^8 M_\odot$ a scale radius of $r_s = 625$ pc at $t = 450$ Myr. The grey bands shows the 1$\sigma$ scatter of the N-body simulations and the black curve shows the curve predicted by the analytic model.  Note that the scales on the $y$ axes are different from \protect\figref{fig:1e7p5_shape} and \protect\figref{fig:1e7_shape}.}
\label{fig:1e8_shape}
\end{figure}

\begin{figure}
\centering
\includegraphics[width=0.5\textwidth]{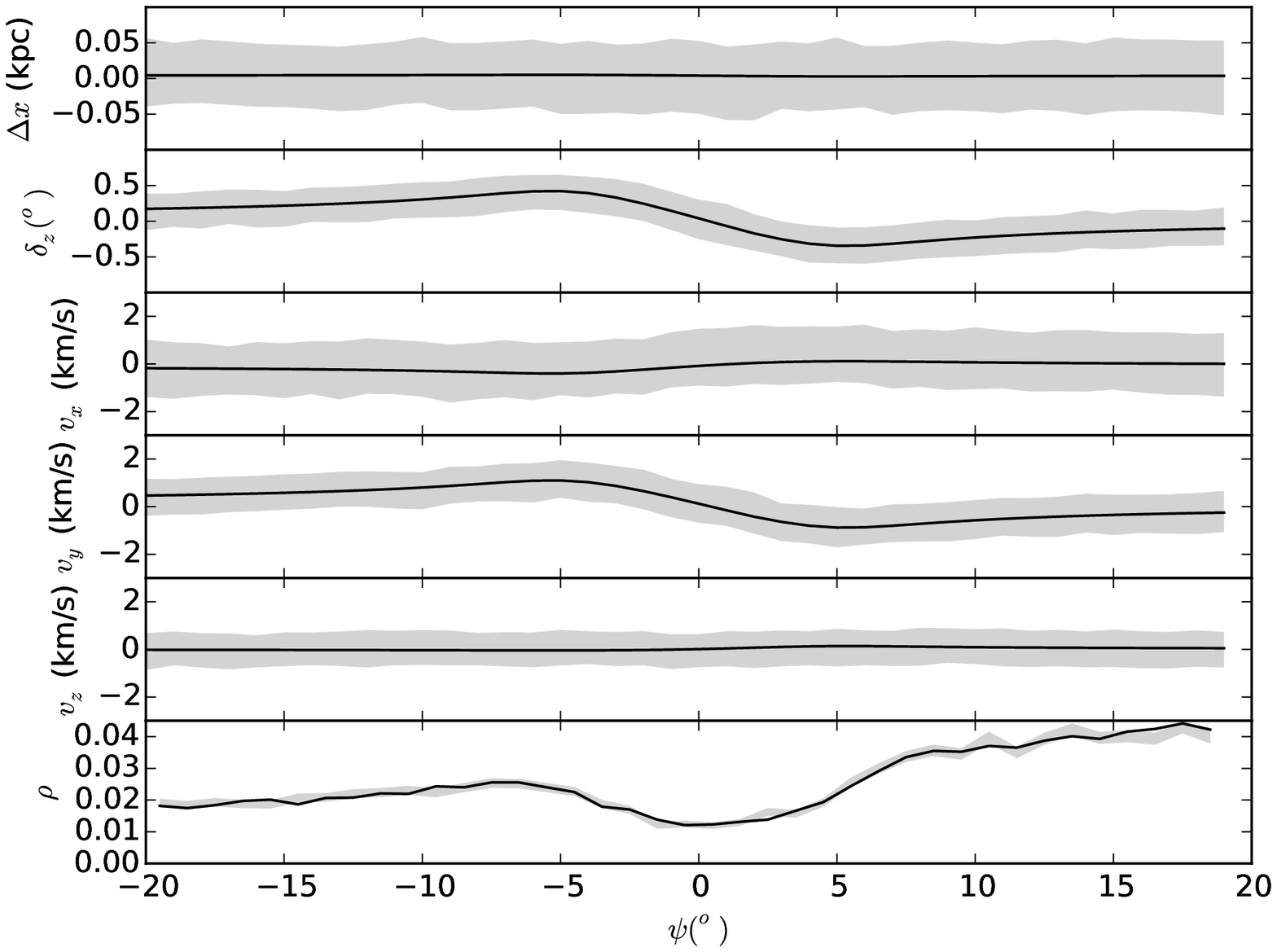}
\caption{Gap observables for a flyby of a Plummer sphere with a mass of $10^{7.5} M_\odot$ a scale radius of $r_s = 395$ pc at $t=450$ Myr. The grey bands shows the 1$\sigma$ scatter of the N-body simulations and the black curve shows the curve predicted by the analytic model. Note that the scales on the $y$ axes are different from \protect\figref{fig:1e8_shape} and \protect\figref{fig:1e7_shape}.}
\label{fig:1e7p5_shape}
\end{figure}

\begin{figure}
\centering
\includegraphics[width=0.5\textwidth]{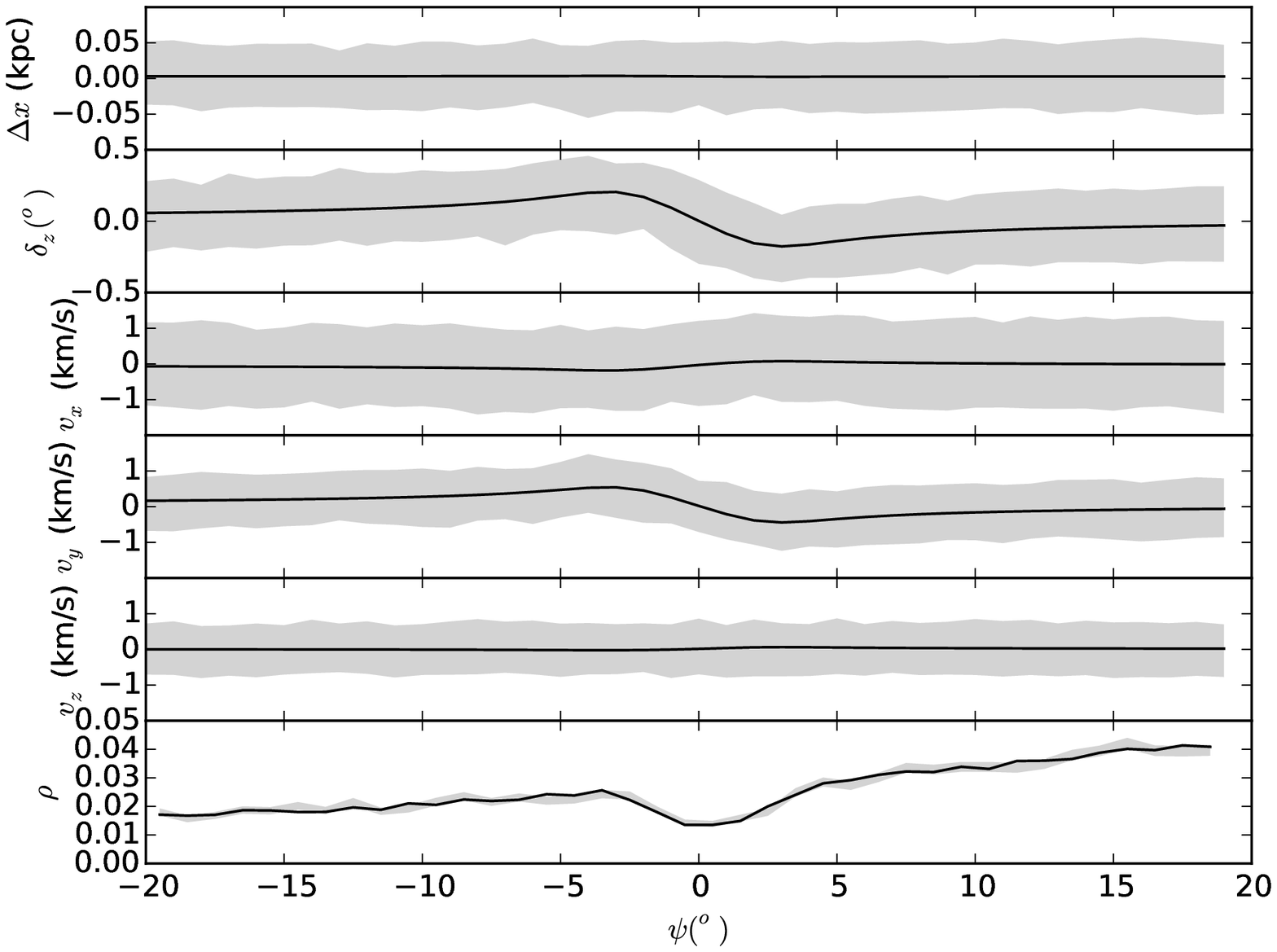}
\caption{Gap observables for a flyby of a Plummer sphere with a mass of $10^7 M_\odot$ a scale radius of $r_s = 250$ pc at $t=450$ Myr. The grey bands shows the 1$\sigma$ scatter of the N-body simulations and the black curve shows the curve predicted by the analytic model. Note that the scales on the $y$ axes are different from \protect\figref{fig:1e8_shape} and \protect\figref{fig:1e7p5_shape}.}
\label{fig:1e7_shape}
\end{figure}

In \Figref{fig:1e8_shape,fig:1e7p5_shape,fig:1e7_shape} we show the six observables for each of the three instances of the perturbed stream at these times. Note that the scales on the axes are different in each of the figures. The grey bands in these figures show the width of the stream and the black curves show the prediction from the model for each setup. We stress that the black curves are not fits to the data but rather a prediction of the model using the true parameters. While the model agrees well with the simulations, it does not capture the entire behavior. For example, if we look at the $v_x$ panel in \Figref{fig:1e8_shape} or \Figref{fig:1e7p5_shape}, we see that the model does not pass through the center of the grey band. This slight disagreement is important as it will result in biases when we fit our model to the simulations as demonstrated in \Secref{sec:basic_inference} and \Secref{sec:error_inference}.

In these simulations, we have focused on direct impacts which will produce the largest signatures. The analytic results of \Secref{sec:observables} allow us to explore how large the signatures will be if we change the impact parameter, the mass of the subhalo, or the velocity of the flyby. The size of the perturbation is governed by the size of the velocity kick which can also be estimated qualitatively by taking the typical acceleration during the flyby, $\sim \frac{GM}{b^2+r_s^2}$, multiplied by the duration of the flyby, $\sim \frac{\sqrt{b^2+r_s^2}}{w}$, to give a velocity kick of $\sim  \frac{GM}{w \sqrt{b^2 + r_s^2}}$. As we showed in Figure 3 of Paper I, the maximum kick in the $y$ direction is given by $\frac{GM w_\perp}{w^2 \sqrt{b^2 + r_s^2}}$ so the qualitative approach almost gives the correct behavior. Thus we see that as we increase the mass of the subhalo, we will get a larger signature. Likewise, if we increase the velocity of the flyby, the impact parameter, or the subhalo's size, we will get a smaller signature. The maximum size of the velocity kicks in the $x$ and $z$ can also be computed using \eqref{eq:deltav} and exhibit roughly the same behavior although the dependence on the velocities, $r_s$, and $b$ is more complicated.

\subsection{Inference} \label{sec:MCMC_no_error}

We will now infer the properties of the subhalo without accounting for any observational error. This showcases how well this method works in the best case scenario. In \Secref{sec:error_inference} we will do the same exercise accounting for observational errors for both contemporary and future surveys. We use emcee \citep{emcee} to explore the likelihood landscape of our models.

The MCMC samples the seven model parameters, $M,r_s,w_x,w_y,w_z,b,t$, as well as 6 nuisance parameters for overall shifts in the 6 phase space coordinates. These nuisance parameters are needed to account for possible offsets of the center of the gap due to epicyclic motion. As we can see from \eqref{eq:deltav}, the center of the gap receives no kick along the stream so it has the same orbital period as the unperturbed stream. However, it can receive kicks in the other directions which causes the center of the gap to execute epicycles around its unperturbed position. As a result, when we measure the location of the gap, we do not know the current offset of the center of the gap from the unperturbed stream and we need to marginalize over this. 

The model described in \Secref{sec:observables} gives the centroid of
the gap in all six observables. The centroid of the gap in the N-body
simulations is found by fitting a Gaussian profile (controlled by 2
parameters describing its center and intrinsic width) to each
observable in 1 degree bins along the stream. This gives the centroid
as well as its error for all observables except the density which is
given Poisson errors. This procedure mimics the process of decoupling
the stream and the background, where, at least for some of the
observables, the separation is done in bins of angle along the stream
and not per star.

With regards to modelling the subhalo-stream encounter itself, the
likelihood for each observable is simply:
\eq{ \ln \mathcal{L} = \sum_i \left( -\frac{\ln(2\pi\sigma_i^2)}{2} - \frac{(o_i-m_i)^2}{2\sigma_i^2} \right),}
where $i$ denotes the angle bins along the stream, $o_i, \sigma_i$ are the centroid and its error in the N-body simulation,  and $m_i$ is the centroid from the model. We assume the observables are independent so the full log-likelihood is the sum of the individual log-likelihoods for each observable.

In the analysis presented here, our priors are as follows. For the
mass, we assume a uniform prior in $\log M$ between a mass of $10^6
M_\odot$ and $10^9 M_\odot$. For the scale radius, we assume a uniform
prior between $1$ pc and $2$ kpc. For the impact parameter, we take a
uniform prior between $-2$ kpc and $2$ kpc. The prior on the velocity
of the subhalo is a Maxwellian distribution with a dispersion, $\sigma$, related to the circular velocity at $10$ kpc, $v_{\rm circ}$, by the relation $\sigma = v_{\rm circ}/\sqrt{3}$. For our setup this gives $\sigma = 97.1$ km/s. This prior is very important since,
as we will see and discuss below, there is a degeneracy between the
mass and velocity of the flyby. For the time since impact, we assume a
uniform prior between $100$ Myr and $2$ Gyr. Next we have priors for
the nuisance parameters. The shift in the angle along the stream is
uniform from $-5$ to $5 \degree$. The shift in the radial distance is
uniform from $-0.2$ kpc to $0.2$ kpc. The shift in all velocity
components is uniform from $-2$ to $2$ km/s.

\begin{figure*}
\centering
\includegraphics[width=0.9\textwidth]{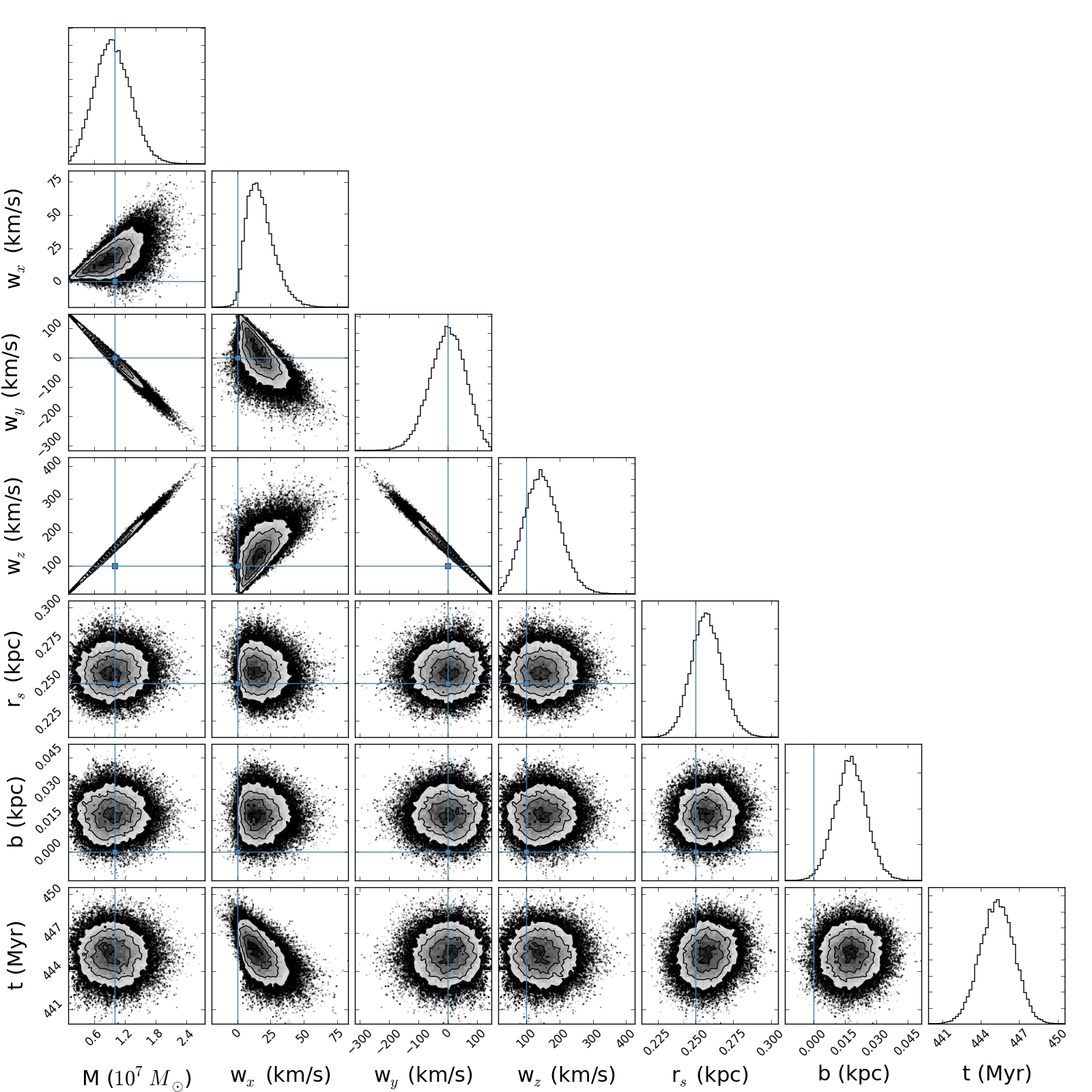}
\caption{Result of MCMC for $10^{7} M_\odot$ subhalo with $r_s = 250$ pc with no errors. Almost all of the subhalo properties are well constrained. However, there is a degeneracy between the mass and the velocity which we will discuss in \protect\Secref{sec:information_content}. }
\label{fig:1e7_MCMC}
\end{figure*}

In \Figref{fig:1e7_MCMC} we show the result of the MCMC modelling for the lowest mass subhalo. We see that without accounting for observational errors, this method can accurately infer the properties of subhalo with a mass of $10^{7} M_\odot$. Almost all of the parameters are well constrained. The only exception is the degeneracy between the mass and the velocity which we will discuss in detail in \Secref{sec:information_content}. This degeneracy is due to the fact that the model cannot infer $M$ and $w$ independently, but rather can only deduce their ratio $M/w$. As a result, the constraint on the mass depends on the prior we place on the subhalo velocity. Note that the likelihood does not peak at the true value for every parameter. In fact, $w_x, b,$ and especially $t$ are discrepant. This is likely due to the fact that our model does not account for all of the features in the stream, such as epicyclic feathering \cite[e.g.][]{kuepper_et_al_2008,kuepper_et_al_2010} and that we assume that the stream is on a circular orbit. While the progenitor itself is on a circular orbit, the stream it generates will be slightly eccentric. For the subhaloes with $10^{7.5} M_\odot$ and $10^8 M_\odot$, the figures look similar albeit with slightly less scatter.

\section{Degeneracy and Information Content} \label{sec:information_content}

As we saw in \Figref{fig:1e7_MCMC}, there is a degeneracy between the mass and velocity of the subhalo. We will now study the set of degeneracies that arise when inferring stream properties.

\subsection{Degeneracy in the velocity kicks} \label{sec:degen_vel}

First, we will study the degeneracy present in the velocity kicks, $\Delta v_x, \Delta v_y, \Delta v_z$, induced by the subhalo's passage. These kicks govern the subsequent evolution of the gap so the best we can do is recover the information content encoded in the kicks. These kicks are given in \eqref{eq:deltav}, which we will re-write as
\eq{ \Delta v_x(\psi_0) &= \frac{A_1(\vec{p})+\psi_0 A_2(\vec{p})}{\psi_0^2+A_3(\vec{p})}, \nln 
\Delta v_y(\psi_0) &= \frac{\psi_0 A_4(\vec{p})}{\psi_0^2+A_3(\vec{p})}, \nln \Delta v_z(\psi_0) &= \frac{A_5(\vec{p}) + \psi_0 A_6(\vec{p})}{\psi_0^2 + A_3(\vec{p})}, \label{eq:vel_kicks_info}}
where we have re-expressed \eqref{eq:deltav} in terms of the six coefficients, $A_i$, which depend only on the parameters $\vec{p} = \{ M,w_x, w_y, w_z, r_s, b\}$, which describe the flyby. For completeness we give the $A_i$ in \eqref{eq:As}. The entire effect of the flyby is encapsulated in these three kicks which are a function of $\psi_0$. If any of these coefficients are different, a different gap will be produced. In the ideal case, we would measure these functions directly and then read off the coefficients. Degeneracies can only arise if there are directions we can move in the 6d parameter space, $\vec{p}$, which do not change any of these coefficients. If we consider a small perturbation in our parameter space, $\delta \vec{p}$, the degeneracies would need to satisfy a system of linear equations:
\eq{ \delta \vec{p} \cdot \vec{\nabla} A_i = 0,}
where the gradient is taken with respect to $\vec{p}$. This is a system of 6 equations with 6 unknowns. In principle, this could have one solution and no degeneracy. However, for generic parameters, these 6 constraints are not linearly independent and give rise to only 5 independent constraints. Therefore, there is a 1d degeneracy which involves a scaling of mass and velocity:

\eq{ M &\rightarrow \lambda M \nln w_x &\rightarrow \lambda w_x \nln w_\parallel &\rightarrow \lambda w_\parallel \nln w_z &\rightarrow \lambda w_z. \label{eq:kick_degen}}
This degeneracy is insurmountable since it does not affect the kicks themselves and hence has no effect on the gap. 

\subsection{Degeneracies in the observables} \label{sec:degen_obs}

We can now apply this same argument to the observables described in \Secref{sec:observables}. We can then combine various subsets of these observations to see what these combinations can constrain. The best we can hope for is to match the ideal case of a 1d degeneracy. The observables have an additional parameter which must be fit: the time since impact, $t$. Thus when considering observables we have a 7d parameter space which we will denote, $\vec{q} = \{ M,w_x, w_y, w_z, r_s, b,t\}$.

\subsubsection*{Gap density}

The first observable is the gap density along the stream. This density is derived from the map between initial and final conditions and is given in \eqref{eq:psi_t}. We re-write this in terms of the notation in \eqref{eq:vel_kicks_info}:
\eq{ \psi(\psi_0,t) = \psi_0 + \frac{B_1(\vec{q}) + \psi_0 B_2(\vec{q})}{\psi_0^2 + B_3(\vec{q})},\label{eq:psi_t_info}}
where the coefficients $B_i$ are only functions of $\vec{q}$ and are independent of $\psi_0$.

\subsubsection*{$x$ vs $\psi_0$}

Next we have the radial distance along the stream, \eqref{eq:x}, which we re-write as
\eq{ \Delta x(\psi_0,t) = \frac{B_4(\vec{q}) + \psi_0 B_5(\vec{q})}{\psi_0^2+B_3(\vec{q})}.}
\subsubsection*{$z$ vs $\psi_0$}

The stream latitude is governed by \eqref{eq:z}, which we re-write as

\eq{ \Delta z(\psi_0,t) = \frac{B_6(\vec{q}) + \psi_0 B_7 (\vec{q})}{\psi_0^2+B_3(\vec{q})} .}

\subsubsection*{$\dot{x}$ vs $\psi_0$}

The radial velocity is given by \eqref{eq:dotx}, which we re-write as
\eq{ \Delta \dot{x}(\psi_0,t) = \frac{B_8(\vec{q}) + \psi_0 B_9 (\vec{q})}{\psi_0^2+B_3(\vec{q})} .}

\subsubsection*{$\dot{y}$ vs $\psi_0$}

The proper motion along the stream, $\dot{\theta}$, is given in \eqref{eq:doty}, which we re-write as
\eq{ \Delta \dot{y}(\psi_0,t) = \frac{B_{10}(\vec{q}) + \psi_0 B_{11} (\vec{q})}{\psi_0^2+B_3(\vec{q})} .}

\subsubsection*{$\dot{z}$ vs $\psi_0$}

The proper motion in the $z$ direction is give in \eqref{eq:dotz}, which we re-write as
\eq{ \Delta \dot{z}(\psi_0,t) = \frac{B_{12}(\vec{q}) + \psi_0 B_{13} (\vec{q})}{\psi_0^2+B_3(\vec{q})} . \label{eq:z_dot_info}}
Note that the denominator is identical for all of these observables. This is because all of these terms are related to the velocity kicks imparted by the subhalo which have the same denominator (see the denominator of \eqref{eq:deltav}). For completeness we give the $B_i$ in \eqref{eq:Bs}.

In the expressions for the observables above, \eqref{eq:psi_t_info} - \eqref{eq:z_dot_info}, there are 13 separate coefficients which govern the observables. If all observables are measured, we can determine the resulting degeneracies in the parameters, $\vec{q}$, by trying to find variations of the parameters, i.e. $\delta \vec{q}$, which do not change the $B_i$:
\eq{ \delta \vec{q} \cdot \vec{\nabla} B_i = 0 , \label{eq:qgradB}}
where the gradient is taken with respect to $\vec{q}$. By carrying out this procedure, we find that these 13 constraints on 7 unknowns give rise to 6 linearly independent constraints, leaving the same degeneracy as above, i.e. \eqref{eq:kick_degen}. This means that if all observables are measured, we can recover the same information as if we measured the kicks themselves.

Note that this result is for a generic value of the parameters. For specific values of the parameters, it is possible to have additional degeneracies if some of the coefficients vanish, i.e. if $B_i = 0$ for some $i$. For example, since the observables oscillate in time, there will be times when some of the observables are flat and contain less information. Another example is if $w_x=0$. In this case, as we can see from the velocity kicks in \eqref{eq:deltav}, $\Delta v_y$ and $\Delta v_z$ only depend on the combination $b^2+r_s^2$ and not $b$ or $r_s$ independently. As a result, if the observables do not constrain (or poorly constrain) the shape in $x$ and $\dot{x}$, we should expect an additional degeneracy to appear between $b$ and $r_s$ where $b^2+r_s^2$ is constant. This is not a true degeneracy since the gap size does not only depend on $b^2+r_s^2$. However, since the leading order term in the gap size only depends on $b^2+r_s^2$, the gap size is not very sensitive to $b$. As a result, $b$ and $r_s$ can be effectively degenerate. We will see an example of this in \Secref{sec:1e7_error_MCMC}.

\subsection{Degeneracies in typical observational setups} \label{sec:degen_var_setup}

Now that we have discussed how to classify the degeneracy when using all observables, we will consider the degeneracies present in common observational setups which only use certain observables. We will repeat the procedure above, using only the coefficients present in the selected observables. 

\subsubsection{Gap density, gap shape in $z$, and radial velocity} \label{sec:degen_z_xdot}

One of the simplest useful setups involves measuring the density of the stream on the sky, the shape of the stream on the sky, and the radial velocity profile along the stream. These three observables depend on the coefficients $B_1, B_2, B_3, B_6, B_7, B_8$, and $B_9$. Using these coefficients, we assemble and solve the system of 7 linear equations as in \eqref{eq:qgradB}. Interestingly, these 7 equations also result in 6 linearly independent constraints on the 7d parameter space, $\delta \vec{q}$, leaving the degeneracy between mass and velocity, i.e. \eqref{eq:kick_degen}. This setup is quite simple since it only requires the radial component of the velocity which is far easier to measure than proper motion. This approach should also allow us to extend the search for gaps out to larger distances since the errors for these observables do not depend as sensitively on distance as the errors in proper motion. 

\subsubsection{Gap density and gap shape  in $z$}

Another approach is to measure the gap density and gap shape on the sky which can be computed using only photometric surveys and do not require any velocity information. These observables depend on the coefficients $B_1, B_2, B_3, B_6$, and $B_7$. We setup and solve the 5 linear equations as in \eqref{eq:qgradB} and find 5 independent constraints, leaving a 2d degeneracy. One of the dimensions of this degeneracy is the mass-velocity degeneracy discussed above. The second degeneracy is more complicated and is between all of the parameters. 

\subsubsection{Gap shape, radial velocity, proper motions}

In this approach, we make use of all of the observables except the density along the stream, i.e. $x$, $z$, $\dot{x}$, $\dot{y}$, and $\dot{z}$ vs $\psi$. This setup makes use of all the coefficients except $B_1$ and $B_2$. We setup and solve the 11 linear equations as in \eqref{eq:qgradB} and find 6 independent constraints leaving the degeneracy between mass and velocity. This is a useful setup since we need to understand the stripping rate of the progenitor as a function of time to fully model the density along the stream. Since this may prove challenging, it is useful to know that the density along the stream is not needed to constrain the subhalo properties. 

In \Secref{sec:1e7_error_MCMC} we will compare the inferred subhalo properties when using these three subsets of observables against the inferrence when using all observables. 

\section{Inference with imperfect data}  \label{sec:error_inference}

Now we will repeat the inference done in \Secref{sec:MCMC_no_error} and account for observational errors. 

\subsection{Observational Errors  for Current and Future Surveys}

We will now assume three different observational setups to estimate the subhalo mass range accessible with this technique. Their details are available in \Tabref{tab:errors}. The first setup we consider follows closely the properties of the SDSS \citep{sdss_dr9} data. Next, we have the ``Super-SDSS'' survey which consists of DES-like \citep{des} photometry, coupled with Gaia \citep{gaia} astrometry for proper motions, and deep multi-object spectroscopy from VLT \citep[e.g.][]{koposov_vlt_kinematics,gaia_eso}, WEAVE \citep{weave}, and 4MOST \citep{4most} for radial velocities. Finally, we have the LSST setup, which is similar to Super-SDSS but uses LSST \citep{lsst_sb_2} photometry and Gaia+LSST for proper motions to extend observations to much fainter stars.

\begin{table*}
\begin{tabular}{|c|c|c|c|c|}
\hline
Setup & $\alpha, \delta$ & RV & Proper Motion & Distance \\ \hline
SDSS & 0.1", $r_{\rm lim} = 22$ & 5 km/s at $r=16$, 20km/s at $r=20$ & None & None  \\
Super-SDSS & 0.1", $r_{\rm lim} = 24$ & 1 km/s at $r=16$, 5 km/s at $r=21$ & Gaia  & DES photometric accuracy \\
LSST & 0.1", $r_{\rm lim} = 27$ & 1 km/s at $r=16$, 5 km/s at $r=21$ & Gaia+LSST & LSST photometric accuracy \\
\hline
\end{tabular}
\caption{Observational uncertainties for contemporary and upcoming surveys. The first column gives the name of the observational setup. The second column, $\alpha, \delta$, gives the observational error and limiting r-band magnitude for determining positions on the sky. The third column, RV, gives the uncertainty in determining the radial velocity as a function of magnitude. We linearly interpolate the errors between the bounds given. The fourth column, Proper Motion, gives the proper motion errors. The Gaia proper motion errors are given in the Astrometric Performance section of \href{http://www.cosmos.esa.int/web/gaia/science-performance}{http://www.cosmos.esa.int/web/gaia/science-performance}. The LSST proper motion errors are given in Table 3.3 of \protect\cite{lsst_sb_2}. The fifth column, Distance, gives the source of distance errors. The distances are calculated using photometric distances as described in \protect\secref{sec:realistic_stars} and distance errors come from the photometric accuracy of each survey. The DES photometric accuracy is discussed in the text. For the LSST photometric accuracy we use the fits given in Table 3.2 of \protect\cite{lsst_sb_2}. }
\label{tab:errors}
\end{table*}

\subsection{Realistic Stellar Observables} \label{sec:realistic_stars}

In order to understand the subhalo mass range which can be probed by these surveys, we need to populate the stream with a realistic distribution of stars. We use the N-body simulations described in \Secref{sec:simulations} which model the globular cluster with identical mass particles. We use PARSEC stellar evolution models \citep{parsec,chen_parsec,tang_parsec} to assign masses and luminosities to each particle. We chose the initial mass function (IMF) from \cite{chabrier_2001} and used a single isochrone with an age of 10 Gyr, and a metallicity of 0.01 $Z_\odot$. We made use of two photometric systems: SDSS $ugriz$ and Gaia $G$. The stellar mass of each particle was sampled from the corresponding present day mass function until the total stellar mass was equal to $10^5 M_\odot$, the mass of the cluster used in the simulation. This resulted in approximately 360,000 N-body particles being tagged as stars. For each star, the magnitude in each band was then linearly interpolated from the tabulated values in the isochrone table. Given the magnitudes $(g, r, G)$ of each particle, observational noise was added to the position, velocity, and magnitudes using the errors and cuts in \Tabref{tab:errors}. We note that the number of stars above the $r$-band cutoff in \Tabref{tab:errors} and within 20 degrees of the gap center increases dramatically as we consider the deeper surveys. The SDSS setup has 1404 stars, the Super-SDSS setup has 4039 stars, and the LSST setup has 10735 stars. We stress that we do not account for contamination from Milky Way stars or from background galaxies.

The relative distances along the stream were calculated using the photometric parallax method. Given the isochrone of the stream, we use the $g-r$ color to infer the r-band absolute magnitude. The apparent $r$-band magnitude combined with the absolute magnitude gives the distance modulus and hence the distance. We propagate the errors from the photometric accuracy for each color and use the slope of the isochrone to give the error in absolute magnitude. We do not account for uncertainties in the isochrone itself as we are primarily interested in relative distance changes across the gap. For the SDSS setup, we chose not to use any distance measurements since the distance errors are too large to give useful constraints. For the Super-SDSS setup, we use the expected photometric accuracy of DES. To estimate the single epoch photometric accuracy, we use the photometric catalog described in \cite{koposov_des} which is derived from publicly available images from the first year of DES. We assume that all five epochs of DES will have a similar accuracy to get a final photometric accuracy which is $\sqrt{5}$ smaller than the single epoch accuracy. For completeness, we list our expected five epoch DES $r$-band magnitude errors: 0.0005, 0.0096, 0.0019, 0.0038, 0.0079, 0.017, 0.039, 0.091 for $r$-band magnitudes of 17,18,19,20,21,22,23,24 respectively. The $g$-band magnitude errors are similar to these. For the LSST setup, we use the LSST photometric accuracies which are reported in \protect\cite{lsst_sb_2}. 

\subsection{MCMC Details}

The details of the MCMC are almost identical to those presented in \Secref{sec:MCMC_no_error}. The only difference is that the observables for each star are now scattered by the appropriate observational error and this is accounted for when we compute the centroid and its error for each observable along the stream. 

\begin{figure}
\centering
\includegraphics[width=0.5\textwidth]{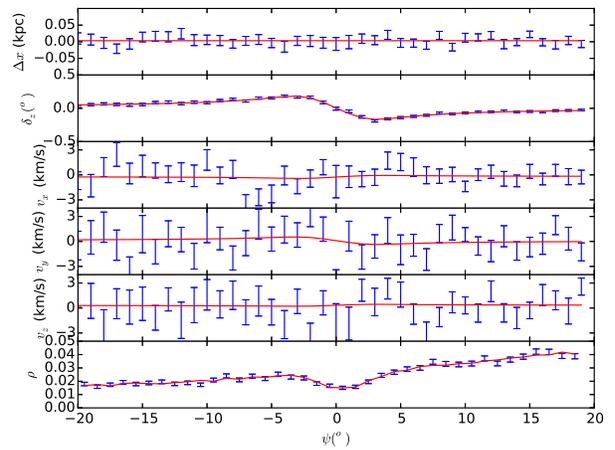}
\caption{Observables for $10^7 M_\odot$ subhalo with LSST errors. The red curves show the maximum likelihood result and the blue error bars show the $1\sigma$ errors of the mock observation.}
\label{fig:observables_1e7_LSST}
\end{figure}

\begin{figure}
\centering
\includegraphics[width=0.5\textwidth]{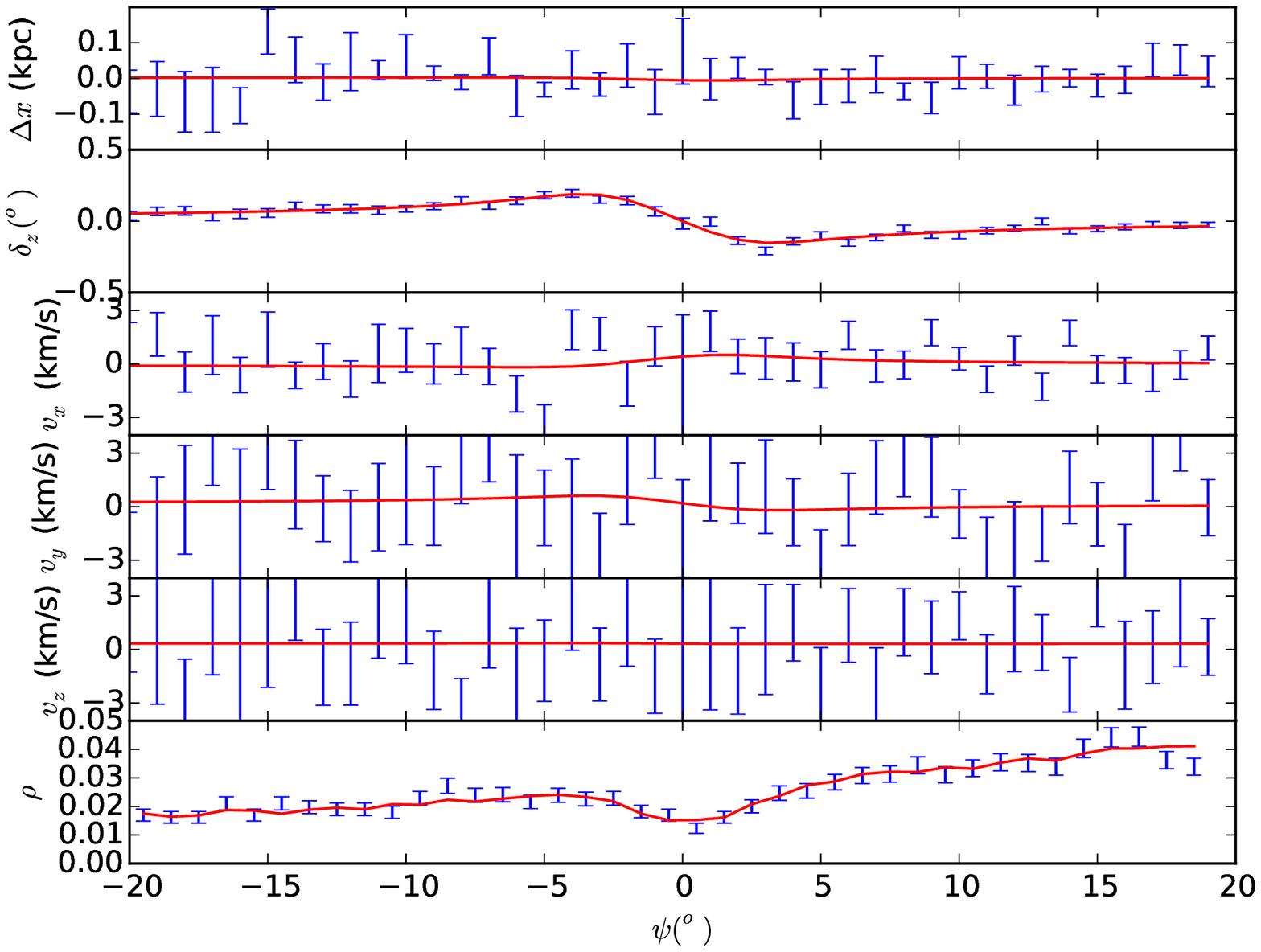}
\caption{Observables for $10^7 M_\odot$ subhalo with Super-SDSS errors. The red curves show the maximum likelihood result and the blue error bars show the $1\sigma$ errors of the mock observation.}
\label{fig:observables_1e7_super_SDSS}
\end{figure}

\begin{figure}
\centering
\includegraphics[width=0.5\textwidth]{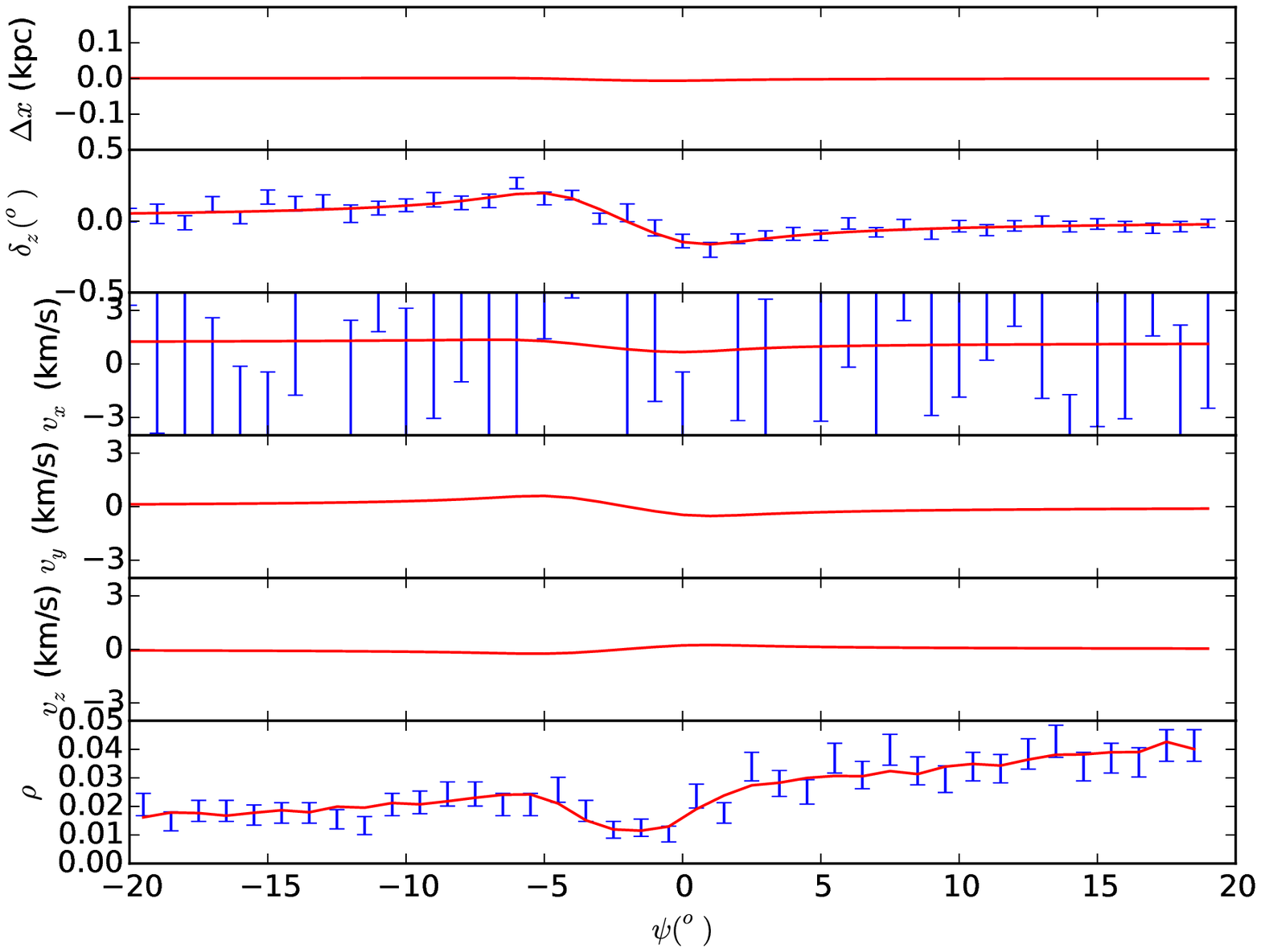}
\caption{Observables for $10^7 M_\odot$ subhalo with SDSS errors. The red curves show the maximum likelihood result and the blue error bars show the $1\sigma$ errors of the mock observation. Note that there are no error bars for $\Delta x, v_y,$ and $v_z$ since these are not used for fitting.}
\label{fig:observables_1e7_SDSS}
\end{figure}

In \Figref{fig:observables_1e7_LSST,fig:observables_1e7_super_SDSS,fig:observables_1e7_SDSS} we show the six observables and their $1 \sigma$ error bars for the $10^7 M_\odot$ subhalo for the three different observational setups: LSST, Super-SDSS, and SDSS. We note that the error bars in these figures are the errors on the mean within each angular bin and are therefore significantly smaller than the error on any individual measurement. We see that, as expected, the shape of the stream on the sky and the density have the smallest error bars, followed by the distance along the stream, and then the radial velocity. We see that the error bars for the Super-SDSS setup are already quite constraining. As we move from Super-SDSS to LSST, the main improvement is the superior photometry and depth (i.e. number of stars) which in combination give smaller error bars in the shape of the stream gap on the sky as well as its density. There is also a noticeable improvement in the proper motion measurements. 

\subsection{Inference} \label{sec:1e7_error_MCMC}

\begin{figure*}
\centering
\includegraphics[width=0.9\textwidth]{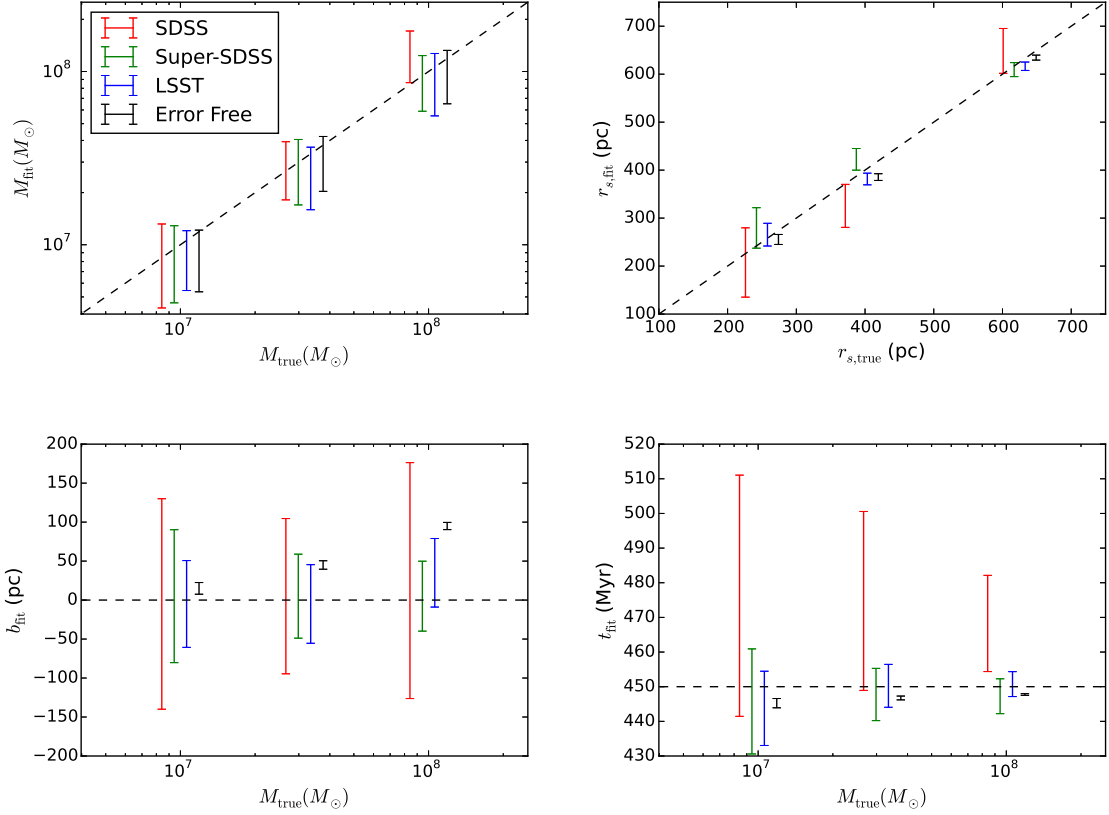}
\caption{Comparison of true and inferred properties for all setups. The error bars show the 1$\sigma$ scatter around the maximum likelihood value. Note that the error bars for each observational setup are shifted slightly along the $x$ axis to allow comparison. On the top left panel we show the true subhalo mass versus the inferred mass. On the top right panel we show the true subhalo scale radius versus the inferred scale radius. On the bottom left panel we show the true mass versus the inferred impact parameter. On the bottom right panel we show the true mass versus the time since impact. For each subhalo setup, we show the constraints for the four observational setups. From left to right, we show the SDSS errors, the Super-SDSS errors, the LSST errors, and the setup with no observational errors. Note that the error free setup shows some bias in $b$ and $t$. This is because our model does not perfectly match the data as described in \protect\Secref{sec:simulations}.}
\label{fig:panel_comparison}
\end{figure*}

\Figref{fig:panel_comparison} shows the 1$\sigma$ constraints on $M, r_s, b,$ and $t$ for each subhalo and each observational setup. For $r_s,b,$ and $t$ we see that the constraints get tighter as we use higher quality data. The exception is $M$ where the constraints do not improve with better data. This is due to the degeneracy between $M$ and $w$ which means that the constraint on $M$ comes mostly from the prior on the subhalo velocity.

\begin{figure*}
\centering
\includegraphics[width=0.9\textwidth]{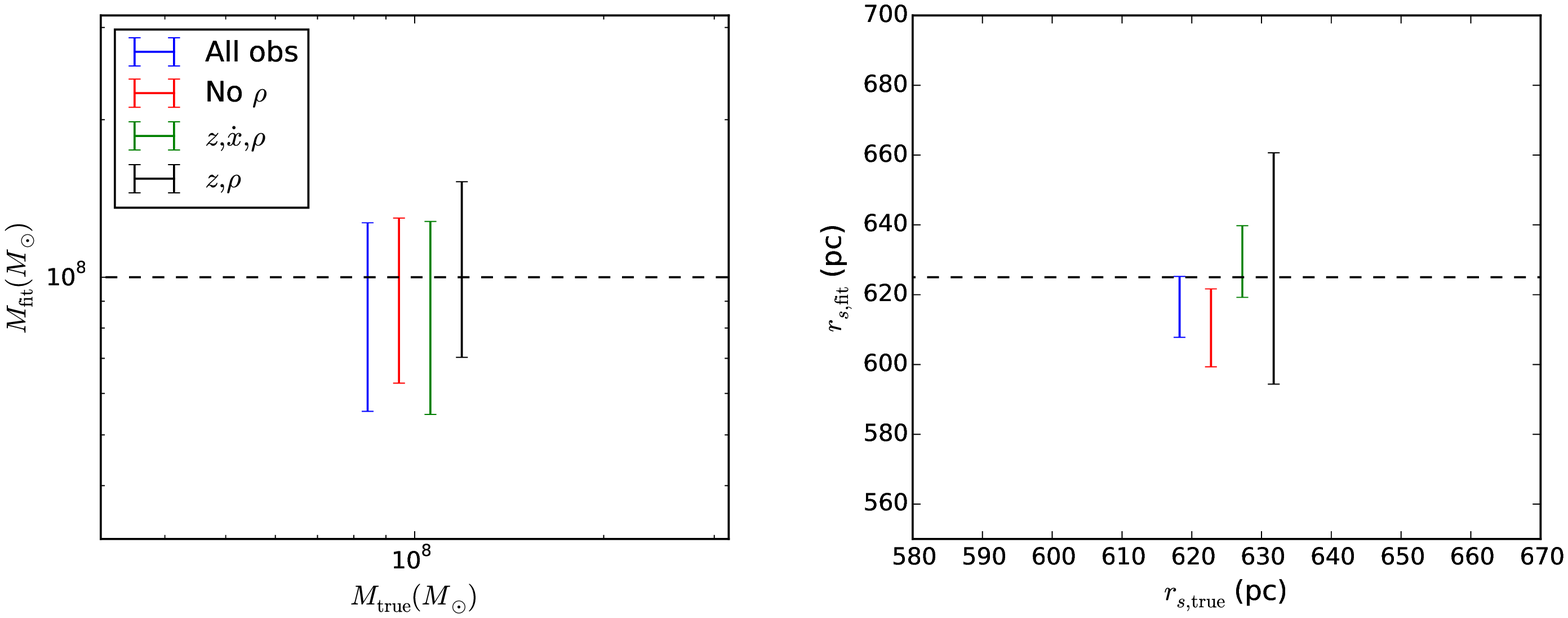}
\caption{Comparison of true and inferred properties for the $10^8 M_\odot$ subhalo with varying amounts of observables using LSST errors. The error bars show the 1$\sigma$ scatter around the maximum likelihood value. Note that the error bars for each observational setup are shifted slightly along the $x$ axis to allow comparison. On the left panel we show the inferred subhalo mass and on the right panel we show the inferred scale radius. We show four setups with varying amounts of information. First, on the left (in blue), we have ``All obs" which uses all of the observables. Moving to the right (in red) we next have ``No $\rho$" which uses all of the observables except for density. Next on the right (in green) is the setup which uses $z,\dot{x},$ and $\rho$. Finally, the rightmost setup (in black) uses only $z$ and $\rho$. We see that the constraints are similar for the first three setups. As we discussed in \protect\Secref{sec:information_content}, these three setups have enough constraining power to constrain all of the observables. If we only measure $z$ and $\rho$ an additional degeneracy is possible and we see a larger error bar in $r_s$.}
\label{fig:panel_comparison_less_info}
\end{figure*}

\Figref{fig:panel_comparison_less_info} gives the 1$\sigma$ constraints on $M$ and $r_s$ for the $10^8 M_\odot$ subhalo using LSST errors. In this figure, we vary the observables used in the MCMC. We see that if we use all of the information, or subsets of the observables which can constrain all of the properties, we obtain results of similar quality. However, if we have too few constraints then the error bars get larger. The three subsets we use here are described in \Secref{sec:degen_var_setup}.

In \Tabref{tab:results} we give the 5-95\% confidence intervals for $M$ and $r_s$ for each subhalo and observational setup. The trends here are identical to those seen in \Figref{fig:panel_comparison}. In \Tabref{tab:results_z_xdot} we show the same intervals for the $10^8 M_\odot$ subhalo when we only use $z$, $\dot{x}$, and density information. As we discussed in \Secref{sec:information_content}, these observables are sufficient to constrain the subhalo properties up to the 1d degeneracy between mass and velocity. As expected, the constraints are similar to when we use all of the information.

\begin{table*}
\begin{tabular}{ c c c c}
\hline
Survey & $M=10^7 M_\odot$, $r_s = 250$ pc & $M= 10^{7.5} M_\odot$, $r_s = 395.3$ pc & $M=10^8 M_\odot$, $r_s = 625$ pc \\ \hline
SDSS &  $M \in [4.49 \times 10^6 M_\odot, 1.94 \times 10^7 M_\odot]$ & $M \in [1.27 \times 10^7 M_\odot, 4.66 \times 10^7 M_\odot]$ & $M \in [6.76 \times 10^7 M_\odot, 2.05 \times 10^8 M_\odot]$ \\
&  $r_s \in [53{\rm pc}, 300{\rm pc}]$&$r_s \in [193{\rm pc}, 358{\rm pc}]$ & $r_s \in [495{\rm pc}, 687{\rm pc}]$ \\ \hline 
Super-SDSS & $M \in [4.69\times 10^6 M_\odot, 1.85 \times 10^7 M_\odot]$ & $M \in [1.46\times10^7 M_\odot, 5.22\times 10^7 M_\odot]$ & $ M \in [4.30\times 10^7 M_\odot, 1.49\times 10^8 M_\odot]$\\
& $r_s \in [171{\rm pc},341{\rm pc}]$ & $r_s \in [380 {\rm pc}, 435 {\rm pc}]$ & $r_s \in [580{\rm pc},634 {\rm pc}]$ \\ \hline
LSST & $M \in [5.09\times 10^6 M_\odot, 1.60 \times 10^7 M_\odot]$& $M \in [1.39\times 10^7 M_\odot, 4.73 \times 10^7 M_\odot]$ & $M \in [4.27 \times 10^7 M_\odot, 1.56 \times 10^8 M_\odot]$ \\
& $r_s \in [217{\rm pc},296{\rm pc}]$ & $r_s \in [353 {\rm pc}, 397 {\rm pc}]$ & $r_s \in [603 {\rm pc}, 631 {\rm pc}]$ \\
\hline
\end{tabular}
\caption{MCMC posteriors for various survey configurations. For each survey and subhalo we list the 5-95\% confidence interval of the posterior. Note that the mass errors do not change appreciably as we change the survey type while the scale radius errors show a more dramatic difference. This is due to the degeneracy between mass and velocity as we discussed in \protect\secref{sec:1e7_error_MCMC}.}
\label{tab:results}
\end{table*}

\begin{table}
\begin{tabular}{ c c}
\hline
Survey & $M=10^8 M_\odot$, $r_s = 625$ pc \\ \hline
SDSS & $M \in [6.76 \times 10^7 M_\odot, 2.05 \times 10^8 M_\odot]$ \\
& $r_s \in [495{\rm pc}, 687{\rm pc}]$ \\ \hline 
Super-SDSS & $M \in [4.37 \times 10^7 M_\odot, 1.54 \times 10^8 M_\odot] $ \\
& $r_s \in [571 {\rm pc}, 642 {\rm pc}]$\\ \hline
LSST &  $M \in [4.56 \times 10^7 M_\odot, 1.56 \times 10^8 M_\odot]$ \\
& $r_s \in [608 {\rm pc}, 640 {\rm pc}]$\\
\hline
\end{tabular}
\caption{MCMC posteriors for various survey configurations when only $z, \dot{x},$ and $\rho$ are measured.  For each survey and subhalo we list the 5-95\% confidence interval of the posterior. Since these three observables are as constraining as when we use all observables (see \protect\secref{sec:degen_z_xdot}), these ranges are very similar to those reported in \protect\Tabref{tab:results}. }
\label{tab:results_z_xdot}
\end{table}

\begin{figure*}
\centering
\includegraphics[width=0.9\textwidth]{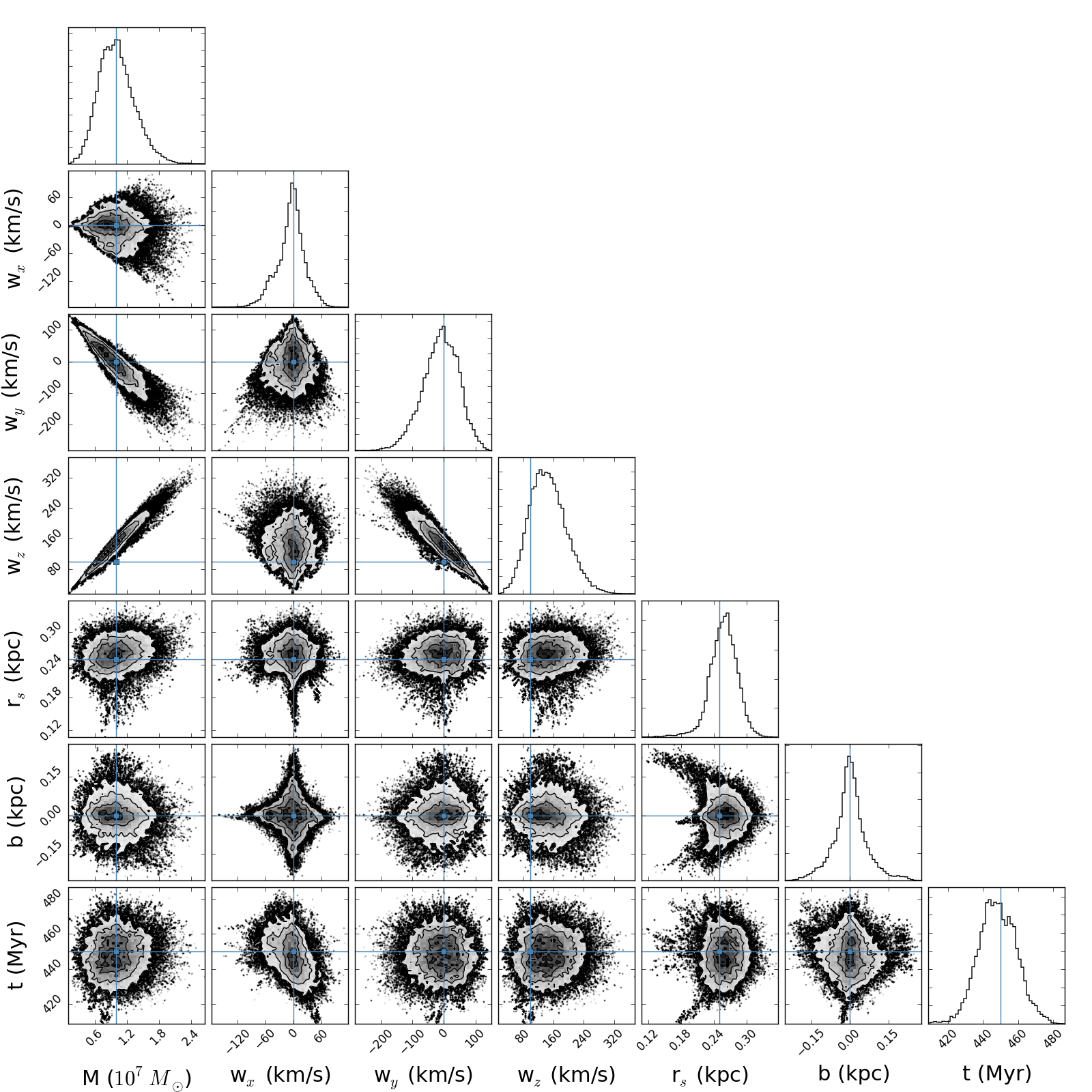}
\caption{Result of MCMC for $10^{7} M_\odot$ subhalo with $r_s = 250$ pc with LSST errors. In addition to the degeneracy between $M$ and $w$, there is a hint of the degeneracy between $b$ and $r_s$ which we describe in the text. This degeneracy is much more apparent in \protect\figref{fig:1e7_SDSS_MCMC} where we use SDSS errors. Apart from these degeneracies, most of the parameters are well constrained in this setup.} 
\label{fig:1e7_LSST_MCMC}
\end{figure*}

\Figref{fig:1e7_LSST_MCMC} presents the posterior distribution for the $10^7 M_\odot$ subhalo when the LSST errors are used. This figure can be compared against \Figref{fig:1e7_MCMC} where we showed the posteriors with no observational error. While there is more scatter for the LSST errors, the overall trend is similar. Most of the parameters are well constrained except for the degeneracy between $M$ and $w$. There is a hint of the degeneracy between $b$ and $r_s$ where $b^2+r_s^2$ is constant and $w_x=0$ which we mentioned in \Secref{sec:information_content}. This can best be seen in the 2d posteriors of $b-w_x$ and $b-r_s$. This degeneracy will be much more vivid when we use SDSS errors.

\begin{figure*}
\centering
\includegraphics[width=0.9\textwidth]{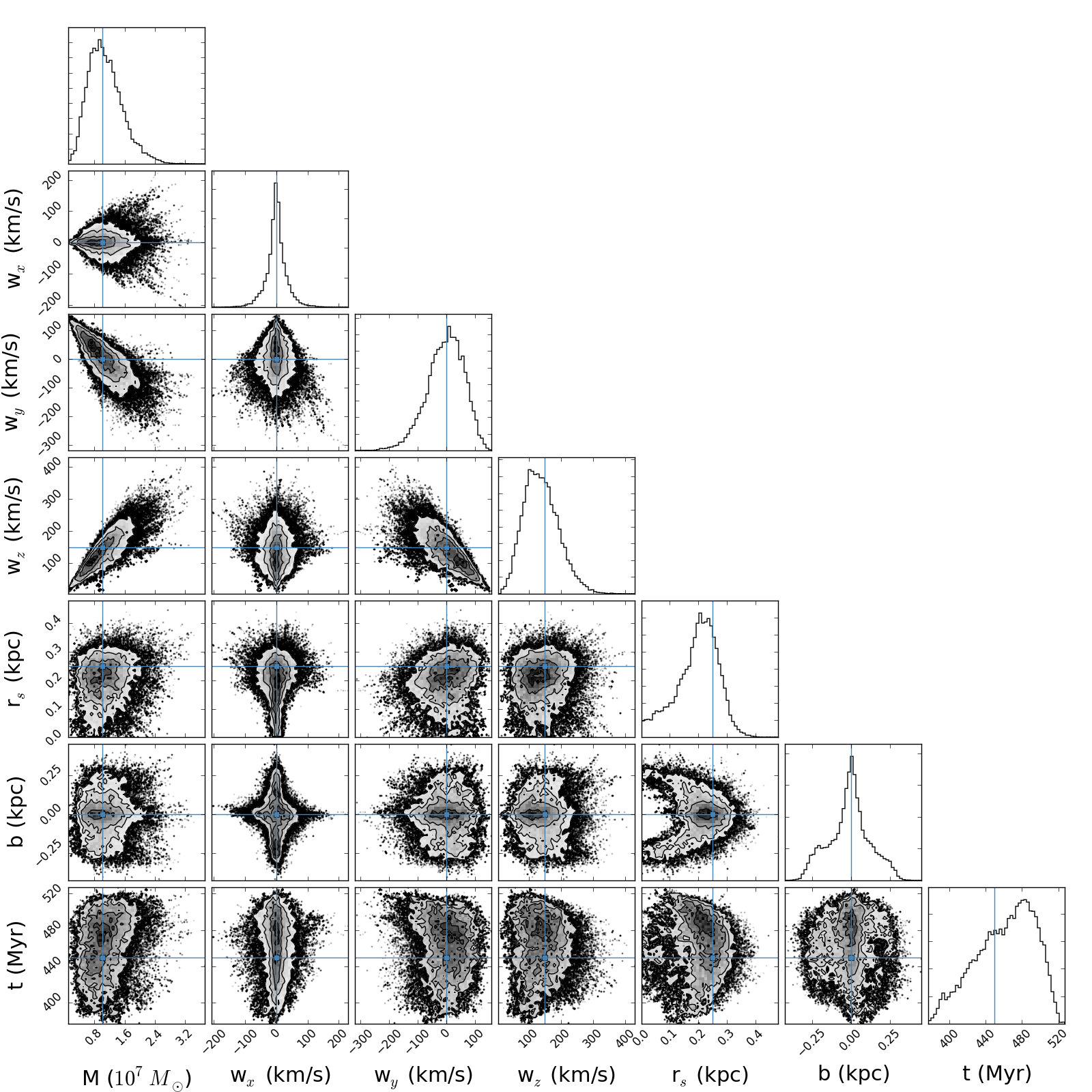}
\caption{Result of MCMC for $10^{7} M_\odot$ subhalo with $r_s = 250$ pc with SDSS errors. In addition to the degeneracy between $M$ and $w$, there is also a degeneracy between $b$ and $r_s$ where $b^2+r_s^2$ is constant. This degeneracy occurs for $w_x=0$ and is discussed in the text.}
\label{fig:1e7_SDSS_MCMC}
\end{figure*}

\Figref{fig:1e7_SDSS_MCMC} demonstrates the posterior distribution for the $10^7 M_\odot$ subhalo when the SDSS errors are used. This setup has significantly more scatter than the LSST setup. The degeneracy between $M$ and $w$ is present as expected. In addition, the degeneracy between $r_s$ and $b$ when $w_x=0$ is now evident. It is especially clear in the 2d posterior of $b-r_s$. This degeneracy is due to the fact that the SDSS setup has no leverage on the distance to the stream and only a modest constraint on the radial velocity. As we discussed in \Secref{sec:information_content}, this allows for an additional degeneracy to appear. Note that this is not a true degeneracy since we have some constraint on the radial velocity and the gap size has a weak dependence on $b$. As a result, the likelihood is still peaked near the true value of $r_s$. 

\section{Mass Range Probed by Current and Future Surveys} \label{sec:mass_range}

In \Secref{sec:error_inference}, we showed the results of a suite of MCMCs where we were able to measure the properties of subhaloes down to $M \sim 10^7 M_\odot$. While this explicit demonstration is useful, we can also use the analytic results of \Secref{sec:observables} to build some intuition as to the constraining power of each observable. We will restrict the analysis to the case of a direct impact, i.e. $b=0$, since these create the largest signature. The detectability of a gap is controlled by the maximum kick imparted by the subhalo. This maximum kick is given by the maxima of \eqref{eq:deltav}:

\eq{ {\rm max} | \Delta v_x | &= \frac{GM}{w r_s} \frac{w_\parallel w_x}{w w_\perp} , \nln 
{\rm max} | \Delta v_y | &= \frac{GM}{w r_s} \frac{w_\perp}{w} , \nln
{\rm max} | \Delta v_z | &= \frac{GM}{w r_s} \frac{w_\parallel w_z}{w w_\perp} . \label{eq:deltav_max}}
Thus we see that the maximum velocity kicks are roughly 
\eq{ \Delta v \approx \frac{GM}{w r_s}. \label{eq:deltav_max_generic} }
We can now plug this expression into those for each of our observables and compare the maximum deviations of the stream against a typical uncertainty on the centroid from the LSST setup. We note that these errors are significantly less than the measurement error for a single star.

\subsection{$x$ vs $\psi$}

The maximum radial deflection is given by the maximum of \eqref{eq:x} which gives
\eq{ \Delta x_{\rm max} = \frac{2 r_0 \Delta v}{\gamma^2 v_y} \Big( 1 + \sqrt{1+\frac{\gamma^2}{4}}\Big) .}
This must be larger than the error in estimating the stream centroid, $\Delta x_{\rm error}$. To determine the mass range accessible with a given error, we plug in the expression for $\Delta v$, i.e. \eqref{eq:deltav_max_generic}, and get
\eq{ \frac{M}{r_s} > \frac{w}{G} \frac{v_y \gamma^2  \Delta x_{\rm error}}{2 r_0 \Big( 1 + \sqrt{1+\frac{\gamma^2}{4}}\Big)}, }
where $w$ is the relative velocity of the flyby so we see that lower velocity flybys allow us to probe a lower mass range. We have re-arranged this expression to isolate the parameters describing the subhalo since these are the most relevant for understanding Galactic substructure. For typical values of $w,v_y \sim 200$ km/s, $r_0 = 10$ kpc, $\gamma^2 = 2$ (i.e. a flat rotation curve), and a radial distance error of $\Delta x_{\rm error} = 15$ pc, we get
\eq{ \frac{M}{r_s} > 6 \times 10^6 \frac{M_\odot}{\rm kpc} \left( \frac{\Delta x_{\rm error}}{15 {\rm pc}} \right) \left( \frac{w}{200 {\rm km/s}} \right) \left( \frac{v_y}{200 {\rm km/s}} \right) \left( \frac{r_0}{10 {\rm kpc}} \right)^{-1}. }
This expression can be used to understand the mass range probed as we change the physical and observational setup. At the edge of this limit, i.e. if $M/r_s = 6 \times 10^6 M_\odot/{\rm kpc}$, we should expect a signal-to-noise ratio of 1. Note that we have not included the $\gamma$ dependence in this expression.

\subsection{$z$ vs $\psi$}

The maximum deflection perpendicular to the plane is given by
\eq{ \Delta \delta_{z, \rm max} = \frac{\Delta v}{v_y}. }
This must be larger than the error in estimating the stream's position on the sky, $\Delta \delta_{z,\rm error}$. This gives 
\eq{ \frac{M}{r_s} > \frac{w }{G} v_y \Delta \delta_{z, \rm error} .}
For the typical values of  $w,v_y \sim 200$ km/s, with the error in the stream latitude is given by the error when estimating the centroid, $\Delta \delta_{z, \rm error} \sim 0.02 \degree$, we get
\eq{ \frac{M}{r_s} > 4 \times 10^6 \frac{M_\odot}{\rm kpc} \left( \frac{\Delta \delta_{z, \rm error}}{0.02 \degree} \right) \left( \frac{w}{200 {\rm km/s}} \right) \left( \frac{v_y}{200 {\rm km/s}} \right). }

\subsection{$\dot{x}$ vs $\psi$}

The maximum deflection in radial velocity is given by
\eq{ \Delta \dot{x}_{\rm max} = \frac{2 \Delta v}{\gamma} \sqrt{1+\frac{\gamma^2}{4}} .}
This must be larger than $\Delta \dot{x}_{\rm error}$, which gives 
\eq{ \frac{M}{r_s} > \frac{w}{G} \frac{\gamma \Delta \dot{x}_{\rm error} }{2 \sqrt{1 + \frac{\gamma^2}{4}} } .}
For typical values with $w \sim 200$ km/s, $\gamma^2 = 2$, and $\Delta \dot{x}_{\rm error} = 1$ km/s, this gives
\eq{ \frac{M}{r_s} > 3 \times 10^7 \frac{M_\odot}{\rm kpc} \left( \frac{\Delta \dot{x}_{\rm error}}{1 {\rm km/s}} \right) \left( \frac{w}{200 {\rm km/s}} \right) .}

\subsection{$\dot{y}$ vs $\psi$}

The maximum deflection in proper motion along the streams's orbit is given by
\eq{ \Delta \dot{y}_{\rm max} = \Delta v \Big( 1 + \frac{2}{\gamma^2}\big(1 + \sqrt{1+\frac{\gamma^2}{4}}\big)\Big) .}
This must be larger than $\Delta \dot{y}_{\rm error}$, giving
\eq{ \frac{M}{r_s} > \frac{w}{G} \frac{\Delta \dot{y}_{\rm error}}{1 + \frac{2}{\gamma^2}\left( 1 + \sqrt{1 + \frac{\gamma^2}{4}} \right) } .}
For typical values with $w \sim 200$ km/s, $\gamma^2 = 2$, and $\Delta \dot{y}_{\rm error} = 2$ km/s at 10 kpc, this gives 
\eq{ \frac{M}{r_s} > 3 \times 10^7 \frac{M_\odot}{\rm kpc} \left( \frac{\Delta \dot{y}_{\rm error}}{2 {\rm km/s}} \right) \left( \frac{w}{200 {\rm km/s}} \right) .}

\subsection{$\dot{z}$ vs $\psi$}

The maximum deflection in the proper motion perpendicular to the orbital plane is given by
\eq{ \Delta \dot{z}_{\rm max} = \Delta v .}
This must be larger than $\Delta \dot{z}_{\rm error}$, which gives
\eq{ \frac{M}{r_s} > \frac{w}{G} \Delta \dot{z}_{\rm error} .}
For typical values of $w \sim 200$ km/s, $\gamma^2 = 2$, and $\Delta \dot{z}_{\rm error} = 2$ km/s at 10 kpc, this gives
\eq{ \frac{M}{r_s} > 9 \times 10^7 \frac{M_\odot}{\rm kpc} \left( \frac{\Delta \dot{z}_{\rm error}}{2 {\rm km/s}} \right) \left( \frac{w}{200 {\rm km/s}} \right).}

By comparing these results, we see that the most discerning observable is the shape of the stream on the sky, followed by the distance to the stream, the radial velocity, and finally the proper motions. The proper motion perpendicular to the orbital plane is less constraining since the velocity oscillations are larger along the orbit. It is not possible to make such a simple estimate for the constraining power of the density along the stream. Note that these formulae can be used to quickly evaluate the mass and size range probed by any survey. 

\section{Discussion} \label{sec:discussion}

The results above show that with current and upcoming surveys, it is
possible to infer the properties of subhaloes down to a mass of
$10^{7} M_\odot$ at a distance of 10 kpc (note, however, that
according to Section~\ref{sec:degen_var_setup}, this measurement can be
extended to much larger distances). In addition, we quantify the
accuracy with which the geometry and the dynamics of the encounter as
well as the mass and the size of the perturber can be
measured. Importantly, we demonstrate that while there exists a
degeneracy between subhalo's mass and velocity, its detrimental effect
can be remedied with the use of reasonable priors. We also presented
expressions for the mass range accessible for a given observational
error in \Secref{sec:mass_range}.

These mass limits should be considered as a best case scenario. In
reality, there will be several additional complications which we will
now discuss. These will introduce additional errors which may raise
the mass limit measurable with the method discussed here. We will also
discuss other causes of substructure in tidal streams and how these
can be distinguished from gaps caused by subhaloes.

\subsection{Discrete Degeneracies}

In \Secref{sec:information_content}, we characterized the degeneracies when inferring subhalo properties. We showed that if we use all of the observables, there is a continuous degeneracy between $M$ and $w$ where $M/w$ is fixed. In addition to this continuous degeneracy, it is possible that there are discrete degeneracies with different points in the parameter space giving rise to the same gap shape. Since the gap shape oscillates in time, if we wait roughly an orbital period, the gap shape will look similar in $z$ vs $\psi$. Of course, the gap will now be larger so we must change the other parameters to get the same gap size and shape. Thus, we might imagine a discrete degeneracy is possible. 

We have searched for this discrete degeneracy by running gradient descents in our negative likelihood space and initiating MCMCs near the location of different likelihood peaks. In the case of no observational errors, we have found no evidence for such degeneracy since the log-likelihoods of the varying peaks differ by more than 1000 from the true peak. As a result, we do not believe this is a true degeneracy. However, if we include observational errors, discrete degeneracies are possible in some cases. For the $10^8 M_\odot$ subhalo and using SDSS observations, we cannot distinguish between solutions which differ by an orbital period. However, if we use Super-SDSS or LSST observations, these solutions differ by roughly 5 and 10 respectively in log-likelihood. For the less massive subhaloes, i.e. $10^{7.5} M_\odot$ and $10^7 M_\odot$, the discrete degeneracies cannot be distinguished even with LSST quality observations. In this work we have only considered one setup: a direct impact on the stream which we observe when the oscillation in the shape on the sky is large. It is not clear how these effective degeneracies will change for a different impact geometry, if we observed the gap at a different time, or if the progenitor is on an eccentric orbit. We will explore this issue in future work.

\subsection{Eccentric orbits}

The analysis done in this work was for progenitors on circular orbits. In reality, progenitors will be on orbits with a range of eccentricities. This will result in the gap stretching and compressing as it passes near pericenter and apocenter respectively. Since analytic orbits are not known for an arbitrary spherical potential, it is not possible to repeat the analysis presented here for eccentric orbits. However, the procedure outlined in this work can be done numerically and the qualitative picture presented here should be the same. In future work, we plan to address this by re-casting this analysis in action-angle formalism. 

\subsection{Uncertainty in the potential}

In the method above, we have assumed perfect knowledge of the potential. Any uncertainty in the potential, and thus the shape of the stream, will complicate the inference of the subhalo properties. However, since the potential is expected to be smooth, its effect on the uncertainty of properties derived from kinks and wiggles over $\sim 10 \degree$ scales in the stream should not be very large. These wiggles can only be created by small scale structure.

\subsection{Oscillations}

One additional complication which we have mentioned above is the fact
that all of these observables oscillate in time. Thus, the quality of
the inference depends on what phase of the oscillation the gap is
in. In the worst case scenario, the observables which have the most
constraining powers (i.e. the shape on the sky or the radial velocity)
will give less information than expected since they are in a phase
with little signal. This will lead to a broader posterior and may
change the results of \Secref{sec:mass_range} so that, in a particular
phase, the proper motion observables are more informative than the
shape of the gap on the sky.

\subsection{NFW Subhaloes}

In this analysis, we have assumed that subhaloes can be modelled as Plummer spheres. In reality, the subhaloes will be closer to NFW profiles \citep{nfw_1997}. Since the kick induced by the passage of an NFW subhalo cannot be computed analytically, this means we will have to compute the kicks numerically. However, since numerics will already be necessary because of the non-circular orbits, this should not introduce much additional complication. Furthermore, since the NFW profile is controlled by two parameters (concentration and virial mass), this more realistic setup will have the same dimensionality as in this work and should exhibit the same qualitative features seen in this work.

\subsection{Epicyclic Feathering}

One physical process can produce density variations in a stream is epicyclic feathering \citep[e.g.][]{kuepper_et_al_2008,kuepper_et_al_2010}. In order to carry out the process described in this work on real streams, we will need to start with a detailed stream model. This model will include these epicyclic feathers so we will be able to determine the likelihood of a gap feature having arisen from the epicyclic feathering. More generally, the epicyclic feathers have different features from the gaps: the epicycles resemble cycloids while the stream near a gap looks like a sinusoid. Thus, with sufficient observations, it should be possible to distinguish the two. This will be easiest if the gap is far from the progenitor, probing an older part of the stream where it is well-ordered. Fortunately, these older parts of the stream are also the most likely to have gaps since they have been around the longest. On a promising note, the N-body simulations in our work naturally included this epicyclic feathering and our model was robust enough to make accurate predictions despite not accounting for these effects. 

\subsection{Gap structure from other substructure}

In addition to dark substructure in the Galaxy, there are also baryonic structures like giant molecular clouds and spiral waves. These can potentially cause confusion but as we have shown in \Secref{sec:basic_inference} and \Secref{sec:error_inference}, the time since impact can be well constrained. This allows us to rewind the position of the gap back to where the impact occurred. If this results in the gap being created in or very close to the disk, baryonic effects should be considered. However, if the impact occurs far from the disk, this should allow us to rule out such effects. In addition, the results of this work show it is possible to get independent constraints on the mass and scale radius. This should give even more evidence to rule out baryonic structures if the gap is created from a large and low-density structure.

\subsection{Searching for Stream Gaps}

The primary signatures of gaps are the oscillations in the stream on the sky outlined in this work and the contrast between the low central density and enhanced density on the edge of the gap discussed in Paper I. The results of this work and the results of Paper I can be used to create match filters to search for these features. However, a simpler approach may be to search for qualitative features in the data. Namely, an underdensity with overdensities around it, and an oscillation in the shape of the stream on the sky. This requires only positions of stars from photometric survey data and does not require any velocities. Once such a feature is found, follow-up would be needed to constrain the radial velocities. As we discussed in \Secref{sec:degen_z_xdot}, these three measurements would be sufficient to constrain the subhalo properties.

\section{Conclusions} \label{sec:conclusions}

In this work, we have shown for the first time, how the properties of
a dark subhalo can be extracted from the measurements of the gap it
creates in a stellar tidal stream. We used the analytic results of
Paper I to forecast the evolution of the 6d structure of the gap shape
arising from a subhalo flyby. We then used these predictions to model gaps
produced in N-body simulations. We carried out inference in an
error-free case as well as for observables with realistic noise for a
range of plausible instrumental setups. The main conclusions of our
study are as follows:

(i) The information content of the subhalo-stream interaction is encoded
in the velocity kicks received by the stream stars.

(ii) Because the velocity kick is an integral of the acceleration
imparted by the subhalo onto stream particles, it depends both on the
strength of the perturber's pull as well as the duration of the
encounter. In other words, there exists a degeneracy between the mass
of the subhalo and the relative velocity of the flyby.

(iii) Using the shape of the gap without observational errors, it is possible to uniquely infer the impact geometry, its
dynamics as well as the age for subhaloes with masses $\geq 10^7 M_\odot$.

(iv) Importantly, it is also possible to independently infer the
properties of the dark perturber, such as its scale length and its mass.

(v) If we account for observational errors from ongoing and planned
surveys, it is possible to infer the impact geometry, its dynamics as
well as the age for subhaloes with masses $\geq 10^7 M_\odot$. For example, for the $10^7 M_\odot$ subhalo with a scale radius of 250 pc, the 1$\sigma$ uncertainty on the measurement of the scale radius, the impact
parameter, the time since flyby decreases from 73pc, 135pc, 34.8 Myr for
SDSS to 24 pc, 56 pc and 10.7 Myr for LSST. 

(vi) Due to the mass-velocity degeneracy mentioned above, while it is
possible to infer the mass of the perturber, the accuracy of the
measurement does not scale simply with the signal-to-noise ratio of
the observables. Yet, in our experiments we found the uncertainty of the perturber mass to be typically
on the order of 0.15 dex.

(vii) Since many of the gap properties oscillate in time, a very
similar gap will be produced by subhalo hits separated by an orbital
period. While these different solutions can be ruled out in the
error-free case, if we add observational errors these multiple
solutions cannot be distinguished for the $10^7 M_\odot$ and $10^{7.5}
M_\odot$ subhaloes.

(viii) Using the analytic model for the stream, we explore the
information contained in several different combinations of
observables. We find that just three measurements, the stream density,
the stream shape on the sky, and the radial velocity along the stream
contain enough information to constrain the subhalo properties. We
believe, therefore, that the stream gap analysis presented here can be
carried out at heliocentric distances much larger than presented here.

The analytic nature of the results in this work allows us to build a
simple intuitive picture of how to infer subhalo properties from
stream gaps. We discussed the difficulties which could arise when
extending this model to gaps created in more realistic streams. This
will require numerical modelling but we expect the same qualitative
picture we have described here. It should be possible to constrain
subhaloes which impact streams on non-circular orbits up to the same
1d degeneracy seen here. In future work, we will explore this by
re-casting this framework in angle-action coordinates. 

Given the promising results of this study, the outlook for dark
subhalo detection in the Milky Way is bright. If Dark Matter exists,
we expect that these elusive clumps will soon be unveiled.

\section{Acknowledgements}

We would like to thank the Streams group at Cambridge for stimulating
discussions and in particular we thank Sergey Koposov for help in
understanding emcee. We thank Jason Sanders and Jo Bovy for useful
discussions and we thank Wyn Evans for comments on an early version of
the manuscript. We note that all of the MCMC triangle figures were
made using triangle.py \citep{triangle} The research leading to these
results has received funding from the European Research Council under
the European Union's Seventh Framework Programme (FP/2007-2013)/ERC
Grant Agreement no. 308024. We thank the referee for their thorough reading of our manuscript and their comments which improved the clarity of this work.

\bibliographystyle{mn2e_long}
\bibliography{citations_is}

\appendix

\section{Coefficients}

For completeness, we list the coefficients $A_i(\vec{p})$ and $B_i(\vec{q})$ which we introduced in \Secref{sec:degen_vel} and \Secref{sec:degen_obs} respectively:

\eq{ A_1(\vec{p}) &= \frac{2 GM b w w_z}{r_0^2 w_\perp^3} , \nln 
 A_2(\vec{p}) &= - \frac{2 G M w_\parallel w_x}{r_0 w_\perp^2 w} , \nln
 A_3(\vec{p}) &= \frac{(b^2+r_s^2) w^2}{r_0^2 w_\perp^2} , \nln 
 A_4(\vec{p}) &= - \frac{2 G M}{w r_0} , \nln
 A_5(\vec{p}) &= - \frac{2 G M b w w_x}{r_0^2 w_\perp^3} ,\nln
 A_6(\vec{p}) &= - \frac{2 G M w_\parallel w_z}{r_0 w_\perp^2 w} . \label{eq:As}}

\eq{ B_1(\vec{q}) &=  \frac{4-\gamma^2}{\gamma^2} \frac{t}{\tau} - \frac{4
\sin\big(\frac{\gamma v_y t}{r_0} \big)}{\gamma^3} \frac{r_0 }{v_y \tau} 
+ \frac{2 \Big( 1 - \cos\big( \frac{\gamma v_y t}{r_0}\big) \Big)}{\gamma^2}
\frac{w_\parallel w_x}{w_\perp^2} \frac{r_0 }{v_y \tau} , \nln
 B_2(\vec{q}) &= -\frac{2 \Big( 1 - \cos\big( \frac{\gamma v_y t}{r_0}\big)
\Big)}{\gamma^2} \frac{b w^2 w_z}{r_0 w_\perp^3} \frac{r_0 }{v_y \tau} , \nln
 B_3(\vec{q}) &= \frac{(b^2+r_s^2) w^2}{r_0^2 w_\perp^2} , \nln
B_4(\vec{q}) &= \frac{2 G M b w_z w}{r_0 v_y w_\perp^3} \frac{\sin\big( \frac{\gamma v_y t}{r_0} \big)}{\gamma} , \nln
B_5(\vec{q}) &= - \frac{4 G M }{v_y w}\frac{\big( 1 - \cos(\frac{\gamma v_y t}{r_0}) \big)}{\gamma^2} - \frac{2 G M w_x w_\parallel }{v_y w_\perp^2 w} \frac{\sin(\frac{\gamma v_y t}{r_0})}{\gamma}, \nln
B_6(\vec{q}) &= - \frac{2 G M b w_x w}{r_0 v_y w_\perp^3}\sin(\frac{\gamma v_y t}{r_0}), \nln
B_7(\vec{q}) &= - \frac{2 G M w_\parallel w_z}{v_y w_\perp^2 w} \sin(\frac{\gamma v_y t}{r_0}) , \nln
B_8(\vec{q}) &= \frac{2 G M b w_z w}{r_0^2 w_\perp^3} \cos(\frac{\gamma v_y t}{r_0}), \nln
B_9(\vec{q}) &= - \frac{2 G M w_x w_\parallel}{r_0 w_\perp^2 w} \cos(\frac{\gamma v_y t}{r_0})- \frac{4 G M}{r_0 w} \frac{\sin(\frac{\gamma v_y t}{r_0})}{\gamma}, \nln
B_{10}(\vec{q}) &= -\frac{2 G M w_z w}{r_0 w_\perp^3} \sin(\frac{\gamma v_y t}{r_0}), \nln
B_{11}(\vec{q}) &= \frac{2 G M \left(w_x w_\parallel \gamma^2 \sin(\frac{\gamma v_y t}{r_0}) - w_\perp^2 \left( - 2 + \gamma^2 + 2\gamma^2 \cos(\frac{\gamma v_y t}{r_0}) \right) \right) }{r_0 w_\perp^2 w \gamma^2} , \nln
B_{12}(\vec{q}) &= - \frac{2 G M b w_x w}{r_0^2 w_\perp^3} \cos(\frac{\gamma v_y t}{r_0}), \nln
B_{13}(\vec{q}) &= - \frac{2 G M w_\parallel w_z}{r_0 w_\perp^2 w} \cos(\frac{\gamma v_y t}{r_0}). \label{eq:Bs} }
where $\tau$ is defined in \eqref{eq:tau}.

\end{document}